\documentclass[journal=jpcbfk,manuscript=article,layout=twocolumn]{achemso}
\pdfoutput=1

\usepackage{epsfig,epsf}
\usepackage{graphicx}
\usepackage{amsmath,amssymb}
\usepackage{abstract}

\newcommand{\oc}{\omega_\text{c}}
\newcommand{\estar}{\epsilon^{*}}
\newcommand{\ead}{e_\text{ad}}
\newcommand{\aw}{a_\text{wrap}}
\newcommand{\kt}{k_\text{B}T}

\author{Teresa Ruiz-Herrero}
\email{teresa.ruiz@uam.es}
\author{Enrique Velasco}
\affiliation[Universidad Aut\'onoma de Madrid]
{Departamento de F\'\i sica Te\'orica de la Materia Condensada, Universidad Aut\'onoma de Madrid, Madrid}
\author{Michael F. Hagan}
\email{hagan@brandeis.edu}
\affiliation[Brandeis University]
{Department of Physics, Brandeis University, Waltham, MA}

\title{Mechanisms of budding of nanoscale particles through lipid bilayers}

\begin{document}

\twocolumn[
\begin{onecolabstract}

  We examine the budding of a nanoscale particle through a lipid bilayer using molecular dynamics simulations, free energy calculations, and an elastic theory, with the aim of determining the extent to which equilibrium elasticity theory can describe the factors that control the mechanism and efficiency of budding. The particle is a smooth sphere which experiences attractive interactions to the lipid head groups. Depending on the parameters, we observe four classes of dynamical trajectories: particle adhesion to the membrane, stalled partially wrapped states, budding followed by scission, and membrane rupture. In most regions of parameter space we find that the elastic theory agrees nearly quantitatively with the simulated phase behavior as a function of adhesion strength, membrane bending rigidity, and particle radius. However, at parameter values near the transition between particle adhesion and budding, we observe long-lived partially wrapped states which are not captured by existing elastic theories. These states could constrain the accessible system parameters for those enveloped viruses or drug delivery vehicles which rely on exo- or endocytosis for membrane transport.

\end{onecolabstract}
]

\vspace{10mm}

\section{Introduction}
 The mechanisms by which nanoscale particles cross cell membranes and the factors that control their uptake are essential questions for cellular physiology and modern biomedicine. Regulating the uptake (endocytosis) of nanoparticles is important for nanomedicine applications and for predicting nanoparticle toxicity \cite{Nel2009, Mitragotri2009, Poland2008}. Similarly, during the replication of many viruses an assembled nucleocapsid buds through the cell membrane, simultaneously exiting the cell and acquiring a membrane coating of host origin. Although endocytosis \cite{Liu2009}, viral budding \cite{Welsch2007,Gladnikoff2009}, and scission of budded viruses \cite{Baumgartel2011} can be actively driven or assisted by cell machinery,  both nanoparticle uptake and at least some aspects of budding of viruses or viral proteins \cite{Helenius1977,Bonsdorff1978} can occur passively (without cell machinery or ATP hydrolysis) \cite{Bihan2009,Hurley2010,Solon2005,Popova2010b}. Furthermore, evidence suggests that some viruses do or can undergo passive budding in vivo (e.g. \cite{Hurley2010, Rossman2010}). It is therefore important to establish the aspects of particle budding which are generic to passive transport and thus underlie all forms of particle uptake or egress. In this paper we use elastic theory, moleculary dynamics (MD) simulations, and free energy calculations to characterize the dynamics and thermodynamics of the process by which a particle adheres to a membrane, is passively engulfed, and then spontaneously separates.

  Previous works first studied the equilibrium configurations of budding through a vesicle or infinite membrane as a function of membrane rigidity, particle size, and membrane-particle adhesion energy using elasticity theory \cite{Deserno2002,Gao2005,Zhang2008}. Subsequent studies used simulations to address further aspects of the problem, including Monte Carlo simulations on a randomly triangulated surface representation of a vesicle \cite{Fosnaric2009} and molecular dynamics (MD) on a coarse-grained lipid model \cite{Li2010}   to investigate wrapping of charged particles, density functional theory to study the relationship between particle hydrophobicity and wrapping \cite{Ginzburg2007}, and dissipative particle dynamics (DPD) to study wrapping of a particle by a inhomogeneous bilayer \cite{Smith2007} and wrapping behavior of ligand-coated nanoparticles  \cite{Yue2011}. Recently, the wrapping  behavior of ellipsoidal particles has been studied via DPD \cite{Yang2011b} and MD simulations \cite{Vacha2011}.

While all of these treatments show that the adhesion energy required for wrapping depends on  particle properties and membrane composition, there has not been a thorough comparison of predictions of elasticity theory with  the results of more sophisticated computational models.
 In this work our primary objective is to understand the extent to which simplified elastic models can describe the thermodynamics and/or dynamics of particle uptake. To focus on aspects generic to all forms of exo- or endocytosis, we consider a minimal model in which the membrane is treated as a bilayer of homogeneous composition and the particle is pre-assembled (as in the case of nanoparticles or, e.g. type-D retroviruses \cite{Garoff1998, Demirov2004}), and is spherically symmetric. Thus, in this work we do not consider the effects of membrane inhomogeneity (i.e. lipid rafts) or the association of viral membrane proteins \cite{Welsch2007, Chan2010}. We compare the predictions of the elastic model \cite{Deserno2002} to results of dynamical simulations and free energy calculations.  We find that the phase behavior predicted by the two descriptions agrees nearly quantitatively in most regions of parameter space, but there are important dynamical differences at parameter values near the transition between no uptake and particle budding. In particular, we identify a partially wrapped state which we show to be metastable.

\section*{Methods}

\subsubsection*{ The Membrane Model }

We model the amphiphilic lipids comprising the membrane with a coarse grained implicit solvent model from Cooke et al \cite{Cooke2005}, in which each amphiphile is represented by one head bead and
two tail beads that interact via repulsive WCA potentials \cite{Weeks1971}, Eq.(1) 

\begin{equation}
V_\text{rep}(r)\!=\!\left\{\begin{array}{cr}
\!4 \epsilon_{0} \! \left[ \left( \frac{b}{r}\right) ^{12}-\left( \frac{b}{r}\right) ^{6}\!+\!\frac{1}{4}\right]\! &\!;r\leq r_\text{c} \\
0 &\! ;r>r_\text{c}
\end{array} \right.
\label{eq:wca}
\end{equation}
with $r_\text{c}=2^{1/6}b$ and $b$ is chosen to ensure an effective cylindrical lipid shape: $b_\text{head-head}=b_\text{head-tail}=0.95\sigma$ and $b_\text{head-tail}=\sigma$, where $\sigma$ will turn out to be the typical distance between beads within a model lipid molecule.

The beads belonging to a given lipid are connected through FENE bonds (Eq.(2)) 
\cite{Grest1986} and the linearity of the molecule is achieved via a harmonic spring with rest
length $4\sigma$ between the first and the third bead, Eq.(3) 
\begin{equation}
 V_\text{bond}(r)=-\frac{1}{2}\kappa_\text{bond} r_\infty^{2} \ln \left[ 1-\left( r/r_{\infty}\right) ^{2}\right]
\label{eq:bond}
\end{equation}
where $r_{\infty}=1.5\sigma$

 \begin{equation}
V_\text{bend}(r)=\frac{1}{2}\kappa_\text{bend}\left( r-4\sigma \right)^{2}
\label{eq:bend}
 \end{equation}

Since this is an implicit solvent model, the hydrophobicity is represented by an attractive interaction, Eq.(4), 
between all tail beads.
\small
\begin{equation}
 V_\text{attr}(r)\!=\!\left\{ \!\begin{array}{ccc}\!
 \!-\epsilon_{0} &;& r<r_\text{c} \\
\!-\!\epsilon_{0}\! \cos^{2} \!\frac{\pi (r-r_\text{c})}{2\omega_\text{c}} \! &\! ;\!&\! r_\text{c}\! \leq \! r \! \leq \! r_\text{c} \!+\! \omega_\text{c} \\
\!0 & ; & r>r_\text{c}+\omega_\text{c}
\end{array} \right.
\label{eq:attr}
\end{equation}
\normalsize
This model allows the formation of bilayers with physical properties such as fluidity, area per molecule and bending rigidity that are easily tuned via
$\omega_{c}$. Moreover, diffusivity within the membrane, density, and bending rigidity are in good agreement with values of these parameters measured for  biological membranes\cite{Cooke2005} (Figure 1) 

\begin{figure}[bt]

\begin{center}
 \includegraphics[width=6.0cm,bb=0 0 771 1053]{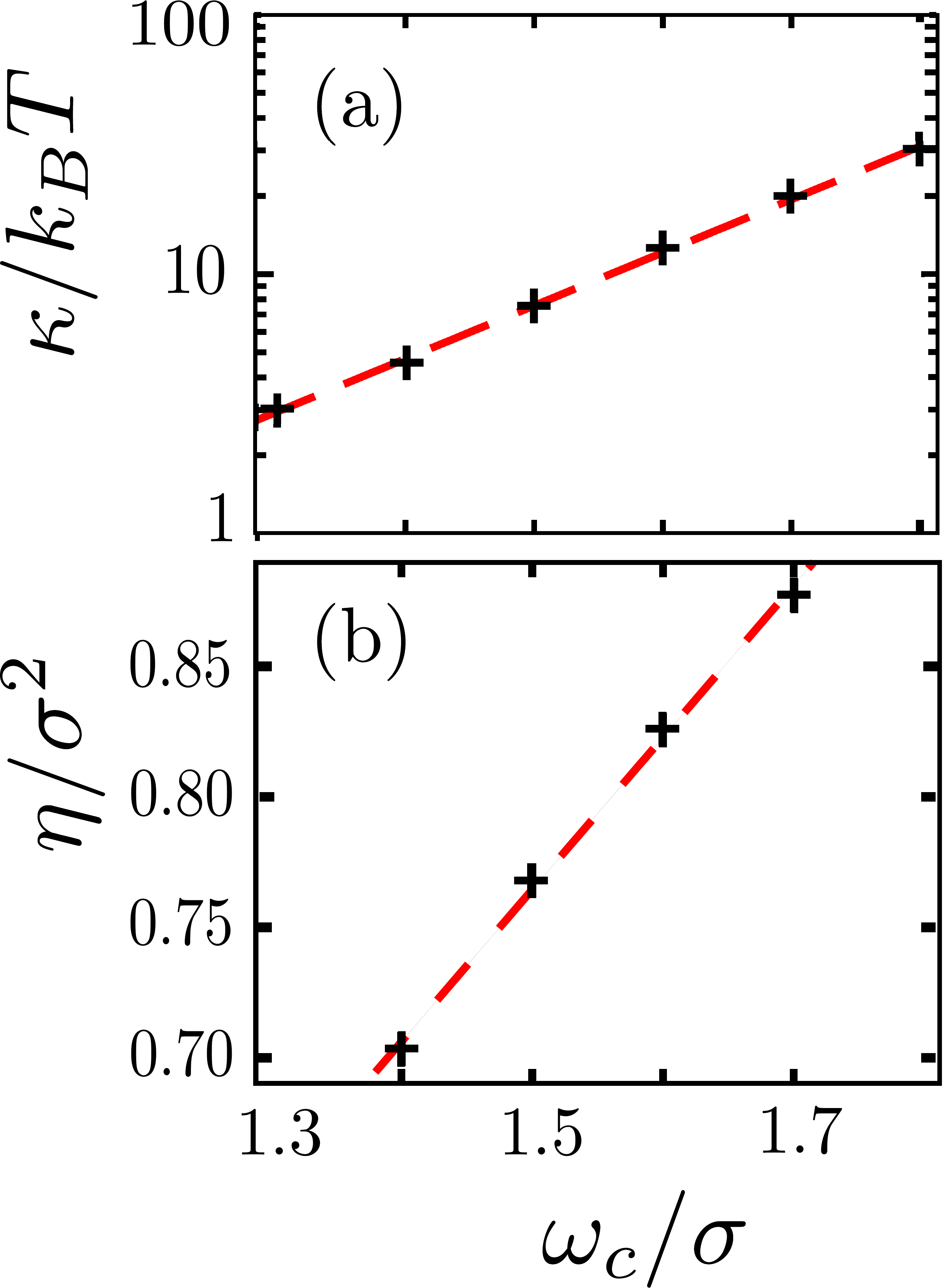}
\end{center}
\caption{\small The properties of the membrane are easily tuned by $\oc$. {\bf (a)} Bending rigidity $\kappa$ (in log scale) and {\bf (b)} areal density $\eta$ of lipids as functions of $\oc$. The values in {\bf (a)} are from Ref. \cite{Cooke2005} and the values in {\bf (b)} were calculated from our simulations.\normalsize}
\label{fig:properties}

\end{figure}

\subsubsection*{Membrane-particle interaction}

As noted in the introduction, the systems we have in mind include synthetic nanoparticles or viral particles which bud through attractive interactions with lipid
 membranes. These interactions can arise in part from electrostatic interactions between charged lipid head groups and charges on the nanoparticle surface or 
capsid exterior (e.g. basic residues on the matrix protein in retroviruses \cite{Welsch2007}). A second source of interaction can be protein mediated, including 
binding of nanoparticle-functionalized ligands to membrane receptors or insertion of hydrophobic tails on capsid proteins into the membrane 
\cite{Garoff2004}.Finally, transmembrane viral `spike' proteins can drive or facilitate budding. Importantly, each of these forms of interactions is short ranged. 
Receptor-ligand and spike protein-virus interactions operate on length scales of \AA{} to nm; similarly, at physiological conditions of 100 mM salt electrostatic 
interactions have a Debye screening length of 1 nm. Thus, to keep our analysis general, we consider a short range attractive interaction between our model 
particle and head groups. In particular, we represent the combination of excluded volume and attractive interactions between the particle and head groups with a 
shifted Lennard Jones potential:

\small
\begin{align}
 V_\text{particle-head}\!=\!\left\{\!\begin{array}{cr}
\! 4 \!\epsilon \!\left[ \!\left( \frac{\sigma}{r\!-\!s}\!\right) ^{12}\!-\!\left( \frac{\sigma}{\!r-\!s}\!\right) ^{6}\!\right]\! - \!V_\text{cut}\! \!&\!;\!r\!<\! r_\text{ph}\!+\!s \\
0 \!& \!;\! r\!\geq\! r_\text{ph}\!+\!s
\end{array} \right.
 \label{eq:vph}
\end{align}
\normalsize
with $\epsilon$ a free parameter that controls the membrane-particle interaction strength, $s=R-\sigma /2$, $R$ the particle radius, \small $V_\text{cut}=4 \epsilon \left[ \left( \frac{\sigma}{r_\text{ph}}\right) ^{12}-\left( \frac{\sigma}{r_\text{ph}}\right) ^{6}\right]$, $r_\text{ph}=3.5\sigma$, and $\epsilon-V_\text{cut}$ \normalsize the depth of the attractive interaction between the particle and the membrane.

The particle experiences only excluded volume interactions with the tail groups, which are modeled with a shifted WCA potential \cite{Weeks1971}, Eq.(6) 
\begin{equation} 	
 V_\text{particle-tail}\!=\!4 \epsilon_{0} \!\left[ \!\left( \!\frac{\sigma}{r\!-\!s}\right) ^{12}\!-\!\left( \frac{\sigma}{r\!-\!s}\!\right) ^{6}\!+\!\frac{1}{4}\right]
 \label{eq:vpt}.
\end{equation}

{\bf Parameters.}
From the phase diagram in \cite{Cooke2005} we set the temperature of our simulations to $k_\text{B}T/\epsilon_{0} =1.1$, which allows for a broad
range of $\omega_\text{c}$ (between $1.3\sigma$ and $1.7\sigma$) within which the membrane is in the fluid state. Furthermore, the bending rigidity was calculated as a function of $\omega_\text{c}$ over this range of values for $k_{B}T/\epsilon_{0} =1.1$ in the same work; results from that reference are shown in Figure 1a. 
Values of the bilayer density calculated in our simulations over a similar range of $\omega_\text{c}$ are shown in Figure 1b. 

The units of energy, length, and time in our simulations are respectively $\epsilon_{0}$, $\sigma$ and $\tau_{0}$. The remaining parameters can be
assigned physical values by setting the system to room temperature, $T=300K$, and noting that the typical
width of a lipid bilayer is around 5 nm, and the mass
of a typical phospholipid is about 660 g/mol. The units of our system can then be assigned
as follows: $\sigma=0.9$ nm, $m_{0}=220$ g/mol, $\epsilon_{0}=3.77 \times 10^{-21} \text{J}=227\text{g}\text{\AA{}}^{2}/\text{ps}^{2}\text{mol}$,
and $\tau_{0}=\sigma\sqrt{m_{0}/\epsilon}=8.86$ ps.

\subsection*{Simulations}

 Molecular dynamics (MD) simulations of budding were performed at constant temperature and pressure using the velocity Verlet algorithm,
with a Langevin thermostat \cite{Frenkel2002a} to maintain constant temperature and a modified Andersen barostat \cite{Kolb1999} to maintain constant membrane tension to represent wrapping by an infinite membrane.
 The time step was $\Delta t=0.01\tau_{0}$, the friction constant was $\gamma = \tau_{0} ^{-1}$, the box friction for the Andersen barostat was $\gamma_\text{box}=2\cdot 10^{-4}$ and the box mass $Q=10^{-5}$ in the system units. The reference pressure, $P_{0}$, is set to 0, to simulate a tensionless membrane. The tension equals the pressure because the normal component to the membrane, the $z$-axis in our case,  is free to fluctuate and does not contribute to the pressure. The $x$ and $y$ components of velocities and positions are rescaled according to the changes in the volume.
In order to simulate an infinite membrane, periodic boundary conditions were employed.

For most simulations the membrane was comprised of $n=21,492$ beads. An initial bilayer configuration was relaxed by MD and then placed normal to the $z$-axis in
a cubic box of side-length $L=63.5 \sigma$. The particle was
introduced in the center of the box with its pole located about 5$\sigma$ below the membrane surface with zero initial velocity.

Since the membrane was kept
tensionless by the barostat, the size of the box decreased during simulations as the particle was wrapped. To ensure that there were no finite size effects, additional sets of simulations were performed, following the same protocol, for membranes with
 $n=48,600$ beads and initial box size of 100x100x60 $\sigma ^{3}$ and with $n=86,400$ beads and initial box size 130x130x60 $\sigma ^{3}$. Except where mentioned otherwise, results are shown for the system with $n=21,492$ beads.

{\bf Free energy calculations.} In addition to performing dynamical simulations of budding, we calculated the potential of mean force as a function of particle penetration using umbrella sampling \cite{Torrie1977}. Simulations were performed in which the system was biased toward particular values of the penetration $p$ by introducing a biasing function $U_\text{bias}(\{\mathbf{r}\})= \frac{1}{2}\kappa_\text{umb}(p(\{\mathbf{r}\})-p_{0})^{2}$. Here $p(\{\mathbf{r}\})$ is the penetration for a configuration $\{\mathbf{r}\}$ and is defined as the distance between the top of the particle and the center of mass of the membrane. A series of windows were performed at different values of $p_0$; for all windows $\kappa_\text{umb}=200\epsilon_0/\sigma^2$. The simulations were started for an unwrapped particle ($p=-6\sigma$), and initial coordinates for each subsequent window were obtained from simulations in the previous one.  Statistics from each window were stitched together and re-weighted to obtain the unbiased free energy using the weighted histogram analysis method \cite{Kumar1992, WHAM}.

\subsection*{Elastic model}
To evaluate the results of the dynamical simulations and free energy calculations, we compare the simulation results to a simplified elastic model for invagination of a particle in a membrane. Our  elastic model closely follows that of Ref. \cite{Deserno2002} but we consider an infinite tensionless membrane rather than a vesicle.

 The total energy of the particle-membrane system arises from the energy of adhesion
between the particle and the membrane ($e_\text{ad}$) and the elastic energy of the membrane ($e_\text{m}$). Following the simulation model, we assume that adhesion is mediated by short-range interactions with energy per area $-\estar$ so that the total energy of adhesion is $\ead=-\estar \aw$ with $\aw$ the area of the membrane in contact with the particle. Note that $\estar$  actually describes a free energy since it includes the effects of counterion dissociation and other entropic factors involved in particle associations, but following Ref. ~\cite{Deserno2002} we refer to it and the elastic terms described next as energies to emphasize that we are neglecting the (small) contribution to the free energy associated with fluctuations around the lowest free energy membrane configuration.

To calculate the elastic contributions to the energy, we consider the Helfrich Hamiltonian for an infinitesimally thin membrane \cite{Helfrich1973}
\begin{equation}
 e_{m}=\int da \left( \sigma_{s} + \frac{\kappa}{2}(2H-C_{0})^{2}+\kappa_\text{G}K \right)
 \label{eq:helfrich}
\end{equation}
where $\sigma_{s}$ is the surface tension, $\kappa$ and $\kappa _{G}$ are the bending rigidity and the Gaussian curvature modulus respectively, H and K are the mean
and Gaussian curvatures, and $C_{0}$ is the spontaneous curvature. Our model membrane is symmetric and tensionless, so $C_{0}$ and $\sigma_{s}$ are 0. We will use this elastic model to describe the budding process up until the point of scission at the neck, and thus the topology of the membrane remains constant. Assuming that the Gaussian curvature modulus $\kappa_\text{G}$ is invariant throughout the membrane, the last term in Eq.(7) 
is constant under the Gauss-Bonnet theorem \cite{DoCarmo1976}. In practice, the properties of the membrane and thus $\kappa_\text{G}$ could change in the vicinity of the adsorbed particle, but this effect contributes a factor proportional to the adhesion area $\aw$ and thus only renormalizes the adhesion free energy $\estar$. The elastic energy for a general configuration of the membrane is then given by
\begin{equation}
e_\text{m}=\int \frac{\kappa}{2}\left( \frac{1}{r_{1}}+\frac{1}{r_{2}}\right) ^{2}da
\label{eq:adhesion}
\end{equation}
with $r_{1}$ and $r_{2}$ the principal radii of curvature.

\begin{figure}[h!]
\begin{center}
 \includegraphics[width=7.5cm,bb=0 0 299 153]{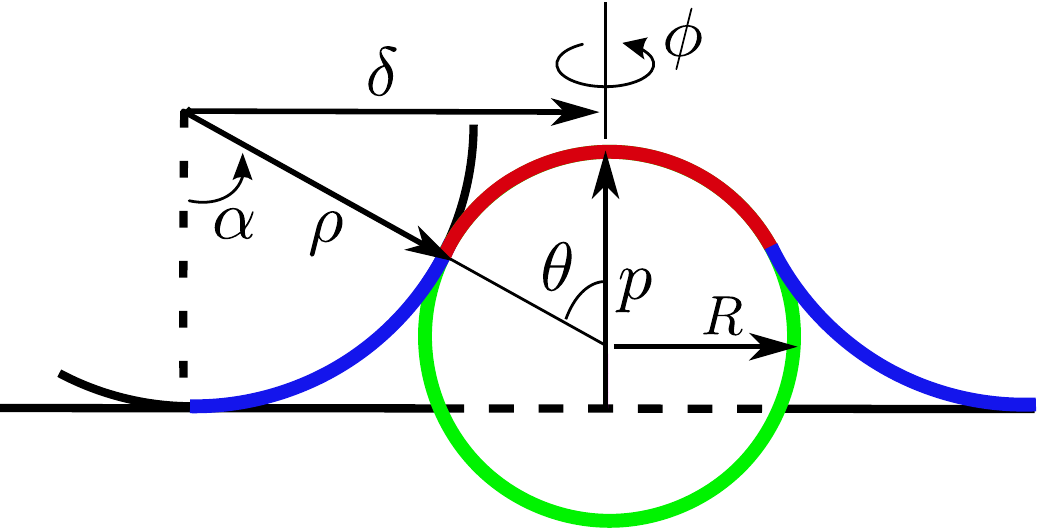}
\end{center}

\caption{\small Cross-section of the 3D geometry used for the elastic model of a membrane wrapping a particle of radius $R$. The particle, depicted as a green sphere, sticks to a section
of the membrane in red with area $\aw$. The surrounding membrane, with area $a_\text{rim}$, drawn in blue, decays toward the flat configuration. The shape of this surrounding membrane is taken
to be a section of a torus for simplicity. $\delta$ and $\rho$ stand for the outer and inner radius of the section of the torus formed by the rim region, and $\alpha$ and $\phi$ represent the polar and azimuthal angle of spherical coordinates. For a given penetration, $p$, there is a wrapping degree, $\theta$, that minimizes the elastic energy.\normalsize}

\label{fig:scheme}
\end{figure}

We follow Deserno et al. \cite{Deserno2002} to assume that the following geometry closely corresponds to the lowest free energy configuration for a partially wrapped particle (Figure 2). 
 There is an area $a_\text{wrap}$ of the membrane tightly adhered to the particle with radius of curvature approximately equal to the particle radius $R$, and a rim area, $a_\text{rim}$, between the point at which the membrane separates from the particle and where it recovers a flat configuration. Because the particle is a featureless sphere, we assume that the lowest free energy configuration is axisymmetric, to give the energy

\begin{equation}
\begin{array}{lc} e=\ead+e_\text{wrap}+e_\text{rim}= \\
 =\!-\! \aw \!\epsilon^{*} \! +\!\kappa \!\left[\!\frac{2\aw}{R^{2}}\!+\!\int \!\frac{da_\text{rim}}{2} \!\left( \!\frac{1}{r_{1}}\!+\!\frac{1}{r_{2}}\right)^{2} \right]
\end{array}
\label{eq:energy1}
\end{equation}

where $a_\text{rim}$ is the area of the rim surrounding the particle, and
$r_{1}$ and $r_{2}$ are the principal radii of curvature in the rim area.

We now recast Eq.(9) 
in terms of two new variables, the latitudinal degree of wrapping $\theta$ and  the penetration $p$, which is the distance the particle travels along the direction normal to the flat membrane, measured from the point at which the surface of the particle first touches the flat membrane.
Note that the theoretical penetration and the one from simulations have different definitions, although qualitatively both describe the system in a similar way. In Figure 2, 
a schematic  of the system is depicted as a 2-D cross-sectional cut, in which the red line represents the section of the membrane bound to the particle and the blue line represents the rim region. As shown in the schematic, we assume that the rim corresponds to a section of a torus (appearing as a circular arc in the 2-D cross-section).  Although this is only one of the multiple shapes the rim can form, it was shown to closely correspond to solutions from a full variational calculation in Ref. \cite{Deserno2002} and  allows us to write the geometric properties of the system as explicit functions of our parameters.
In particular, the radius of the torus depends uniquely on the particle size $R$, the wrapping degree $\theta$ and the penetration $p$;
the area element on a torus and the two principal radii of curvature are \cite{Deserno2002} $\mathrm{d}a_\text{rim}=\rho(\delta-\rho \sin\alpha)d\alpha d\phi$,
$r_{1}={\rho}$ and $r_{2}=-\frac{\delta-\rho \sin\alpha}{\sin\alpha}$, where $\alpha$ and $\phi$ are the polar and azimuthal angle in spherical coordinates.
With the new parametrization, the area of the membrane in contact with the particle for a wrapping degree $\theta$, turns out to be $\aw(\theta)=2\pi R^{2}(1-\cos(\theta))$.

Therefore, the energy of the system can be written in the following way:

\begin{equation}
\begin{array}{lc} e=(-\frac{A\epsilon ^{*}}{2}+4\pi \kappa)(1-\cos\theta) +\\
 
+ \!\pi \kappa\int_{0}^{\theta} \! \rho|\delta\!-\!\rho \sin \alpha | \left(\! \frac{1}{\rho}\!-\!\frac{\sin\alpha}{\delta \!-\!\rho \sin\alpha} \right) ^{2} d\alpha
\end{array}
\label{eq:energy2}
\end{equation}
 where $A=4\pi R^{2}$ is the surface area of the particle.
For a given bending rigidity, particle size and membrane-particle interaction, the energy only depends on the penetration $p$ and the wrapping degree $\theta$. Finally, for each value of $p$ we minimize the energy  Eq.(10) 
with respect to $\theta$ to obtain the membrane configuration and corresponding energy as a function of penetration alone. The results of the minimization are described below in section \textbf{Phase Diagrams}

\section*{Results}
\subsection*{System behavior}

To understand the influence of membrane and particle properties on budding, we began by performing dynamical simulations for a range of particle-membrane interaction strengths, $\epsilon$,
 particle radius $R$, and $\oc$, which controls the areal density of lipids, the bending rigidity $\kappa$, and diffusion rates within the membrane, as
described in section {\bf The membrane model}. Different values of these parameters lead to dramatically different behaviors, as shown in phase diagrams
presented below (Figure 12). 
 First we note that the behaviors can be grouped into four classes, which we illustrate by describing trajectories observed for various
values of $\epsilon$ and constant values of $\kappa=13.9 \kt$ and $R=12\sigma$ (Figure 3). 

\begin{figure}[h!]
\begin{center}
 \includegraphics[width=7.5cm,bb=0 0 1017 626]{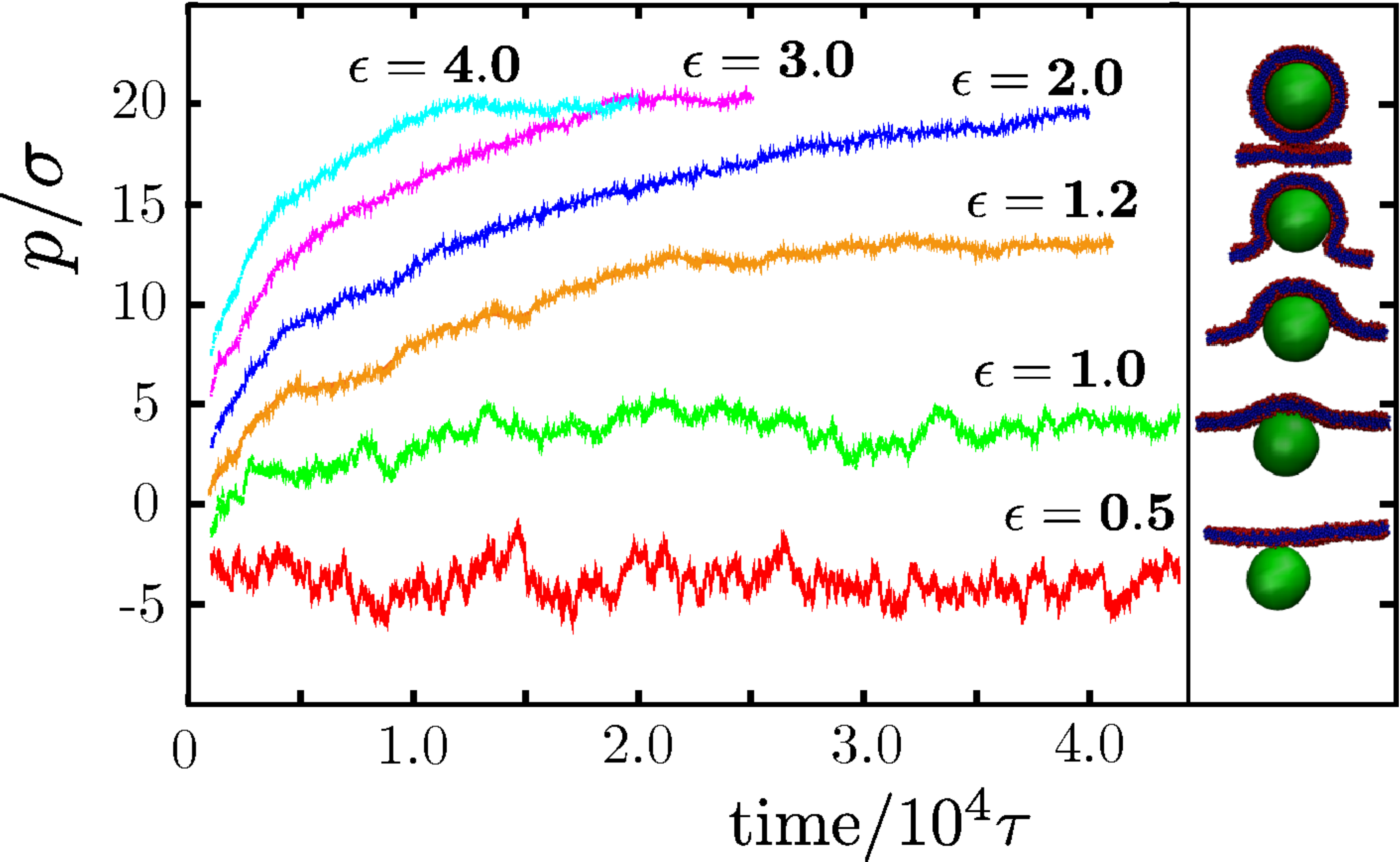}
\end{center}

\caption{\small Particle penetration into the membrane, $p$, as a function of time for molecular dynamics trajectories with different values of the adhesion strength for a system with particle radius $R=12\sigma$ and membrane bending rigidity $\kappa=13.9\kt$. For  $\epsilon=0.5\epsilon_{0}$ (red line) no wrapping occurs. For $\epsilon=1.0\epsilon_{0}$ (green line) and $\epsilon=1.2\epsilon_{0}$ (orange line), budding becomes stalled at a partially wrapped state whose value increases with $\epsilon$. For $\epsilon=2.0\epsilon_{0}$ (blue line), 3.0 (pink line), and 4.0 (cyan line) the particle undergoes complete encapsulation. On the right, slices of the system for corresponding values of $p$ are shown. Images were generated using VMD \cite{Humphrey1996}\normalsize}
\label{fig:md}
\end{figure}

For weak adhesion strengths $\epsilon$, no wrapping occurs; the membrane continues to exhibit only the usual spectrum of thermal fluctuations (Figure 4) 
 after the particle adheres to it, and the penetration oscillates around negative values (Figure 3, 
  case for $\epsilon=0.5\epsilon_{0}$).

\begin{figure}[h!]

\begin{center}
 \includegraphics[width=7.5cm,bb=0 0 782 384]{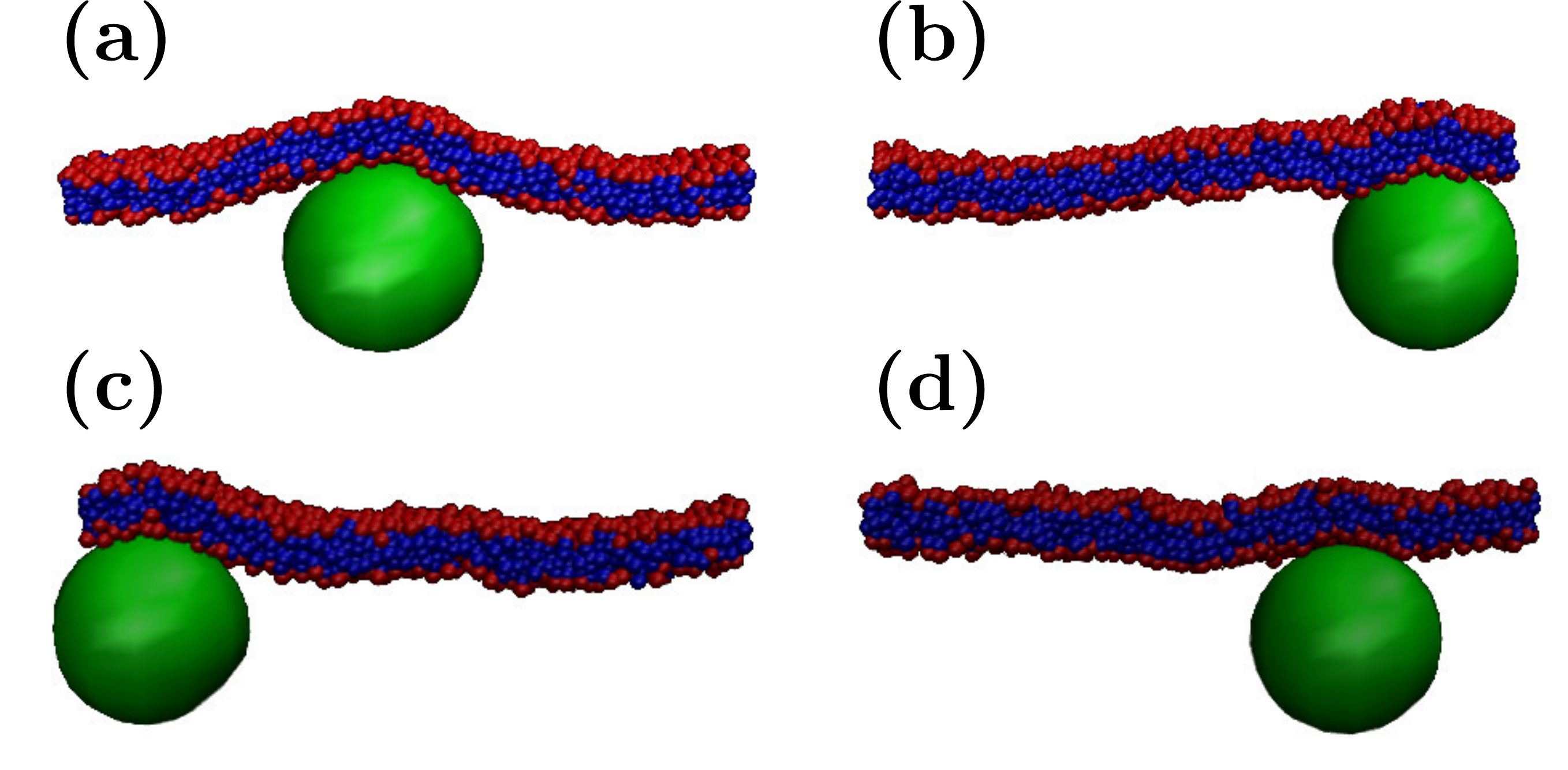}
\end{center}

\caption{\small \textbf{Adhesion without wrapping.} Slices of configurations extracted from  MD simulations with $\epsilon=1\epsilon_{0}$ ,$R=10\sigma$, and $\kappa=13.9\kt$.  Times shown are (a) $t=5\cdot10^{3}\tau_{0}$,
(b) $t=4\cdot10^{4}\tau_{0}$,(c) $t=6\cdot10^{4}\tau_{0}$, (d) $t=1\cdot10^{5}\tau_{0}$. \normalsize }
\label{fig:confnwrap}
\end{figure}

For a narrow intermediate range of $\epsilon$, the particle adheres to the membrane, but wrapping ceases at  a partially wrapped state (Figure 5), 
 after which the degree of particle penetration into the membrane fluctuates around a steady value (Figure 3, 
case for $\epsilon=1.0\epsilon_{0}$). The average value of the penetration remained unchanged  for as long as we simulated (up to $4\cdot 10^{4}\tau_{0}$). The final degree of penetration in this arrested state increases with $\epsilon$, until approximately the point at which the particle is half wrapped (Figure 3, 
 case for $\epsilon=1.2\epsilon_{0}$).
 A further increase in $\epsilon$ results in the next class of trajectories (Figure 3, 
 case for $\epsilon=2.0\epsilon_{0}$ and $\epsilon=3.0\epsilon_{0}$), in which the particle is completely encapsulated (Figure 6). 
In this case, wrapping proceeds steadily until the particle is completely surrounded by membrane except for a narrow neck region (Figure 6d). 
 Wrapping is then completed when a thermal fluctuation causes the neck to break and have its sides fused (Figure 6e), 
 after which the fully wrapped particle diffuses away from the membrane (Figure 6f). 
 Since fusion is a stochastic event, the budding time can be variable and we have observed neck configurations lasting between 500 and 5000 $\tau_{0}$.  The elastic theory predicts that the shape and length of the neck depend on the balance between the adhesion energy and the bending energy with strong adhesion favoring a short neck and large bending energies favoring a long neck. The simulation results are consistent with this prediction; example configurations are shown in Figure 7. 
 The figure shows snapshots from simulations with particle radius $R=6\sigma$, bending rigidity $\kappa=13.9\kt$ and different values of $\epsilon$. A small particle size was chosen for the figure because  the relationship between neck configuration and adhesion energy is most easily visualized when high membrane curvature is required for wrapping. The fact that fusion is accessible within the course of a typical simulation is an interesting contrast between the model studied here and that studied by Smith et al. \cite{Smith2007}, where fusion was observed only for inhomogeneous membranes.

\begin{figure}[h!]

\begin{center}
 \includegraphics[width=7.5cm,bb=0 0 612 290]{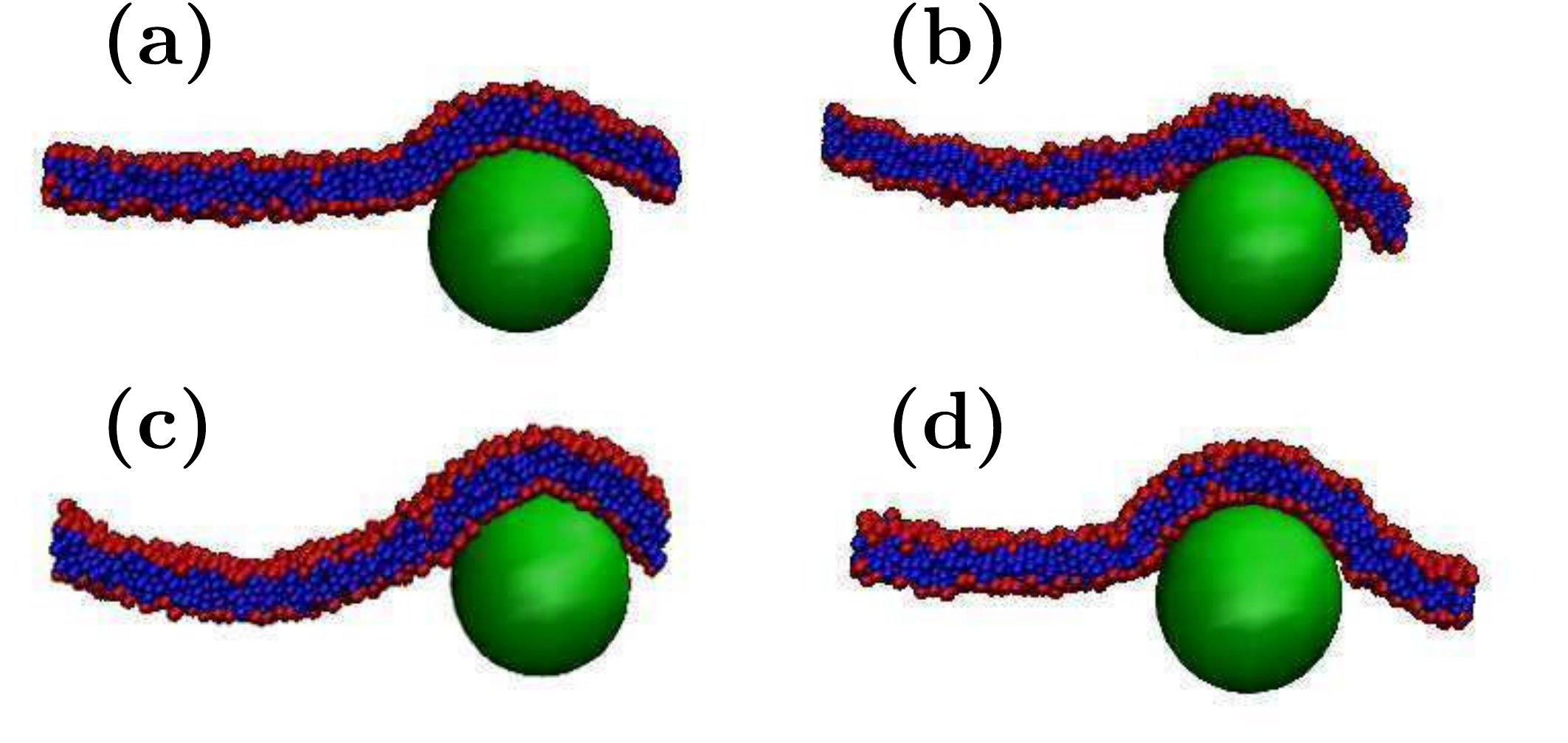}
\end{center}

\caption{\small \textbf{Long-lived partial wrapping.} Slices of configurations extracted from  MD simulations with $\epsilon=1.3\epsilon_{0}$, $R=10\sigma$,  and $\kappa=13.9\kt$. The particle remains partially wrapped for the length of the simulation ($t=4\cdot 10^{4}\tau_0$). (a) $t=5\cdot10^{3}\tau_{0}$,
(b) $t=1.5\cdot10^{4}\tau_{0}$,(c) $t=2.5\cdot10^{4}\tau_{0}$, (d) $t=3\cdot10^{4}\tau_{0}$.\normalsize}
\label{fig:confhwrap}
\end{figure}

\begin{figure}[h!]
\begin{center}
 \includegraphics[width=7.5cm,bb=0 0 982 480]{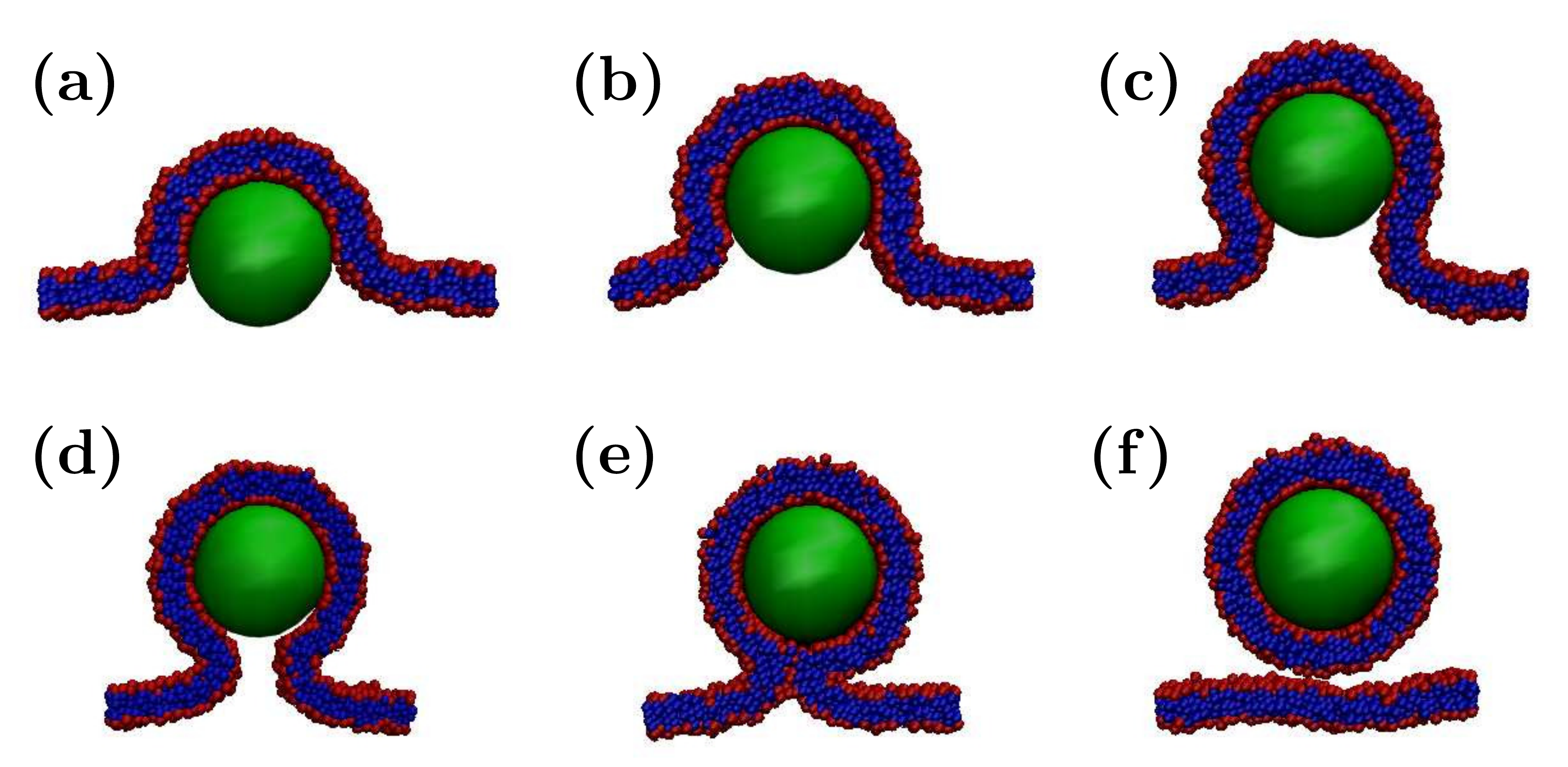}
\end{center}

\caption{\small \textbf{Wrapping.} Slices of configurations at different times extracted from  MD simulations with $\epsilon=3\epsilon_{0},R=10\sigma$ and $\kappa=13.9\kt$.
 The membrane wraps the particle (a-d) until a neck or channel connecting the flat bilayer and the membrane
surrounding the particle forms (e). Thermal fluctuations break this narrow neck, resulting in the encapsulated particle escaping from the membrane (f). Configurations are shown for times (a) $t=5\cdot10^{3}\tau_{0}$,
(b) $t=1\cdot10^{4}\tau_{0}$,(c) $t=1.5\cdot10^{4}\tau_{0}$, (d) $t=1.95\cdot10^{4}\tau_{0}$, (e) $t=2\cdot10^{4}\tau_{0}$,(f) $t=2.5\cdot10^{4}\tau_{0}$.\normalsize}
\label{fig:confwrap}
\end{figure}

For higher values of $\epsilon$ wrapping proceeds extremely rapidly  (Figure 3, 
 case for $\epsilon=4.0\epsilon_{0}$). as there is a strong driving force to increase the number of head-particle interactions (Figure 8). 
As the curvature of the membrane in the vicinity of the wrapping front increases, the membrane structure undergoes ruptures in that region (Figure 8d), 
 and a pore forms in the membrane (Figure 8e). 
 The fully encapsulated particle then diffuses away and the pore heals through thermal motions of the lipids. The formation of a pore during these budding trajectories resembles the process by which a hydrophobic nanoparticle passes through membranes in the simulations described in \cite{Li2008,Yue2011}, but the physical driving forces are different in this case and the pore arises for kinetic reasons. Namely, the collective wrapping process proceeds more slowly than ruptures form in the membrane due to the large driving force to increase particle-head group contacts.

\begin{figure}

\begin{center}
 \includegraphics[width=8.5cm,bb=0 0 825 327]{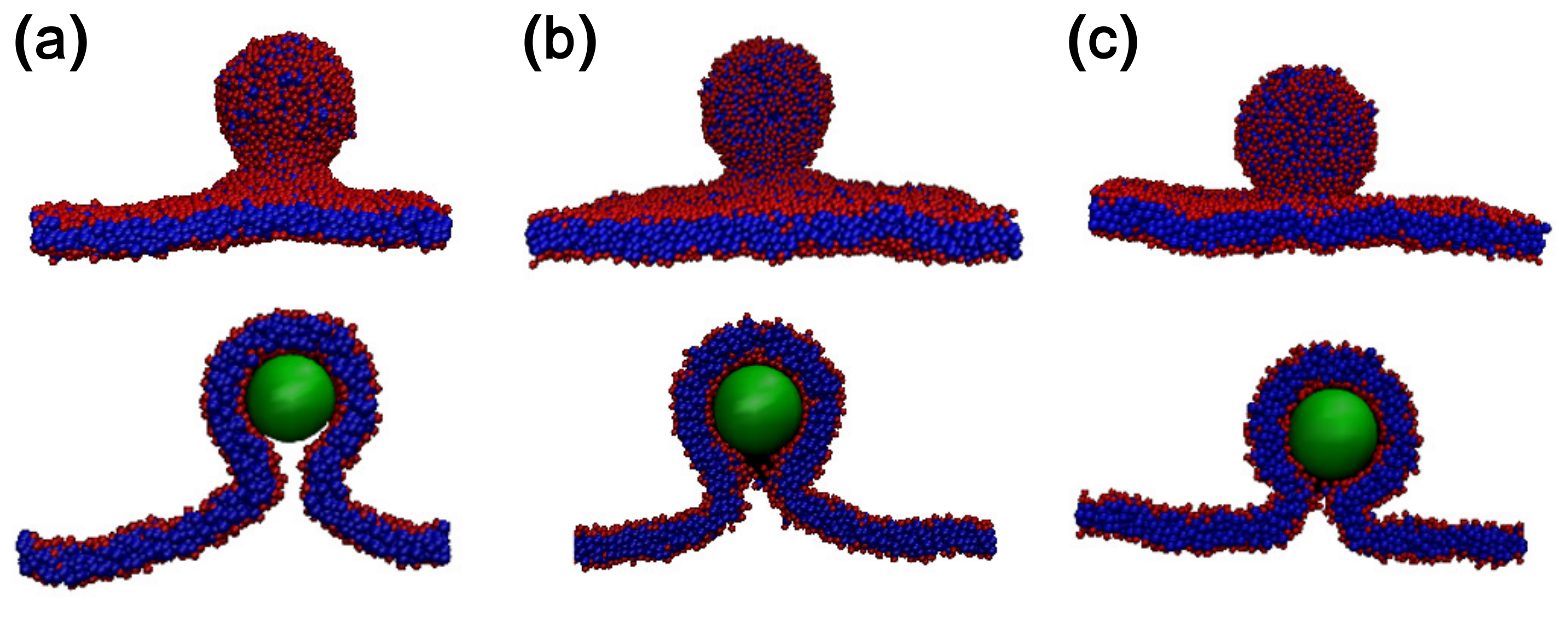}
\end{center}

\caption{\small \textbf{The neck profile depends on the adhesion strength.} Membrane configurations are shown shortly before the completion of budding for $R=6\sigma$, $\kappa=13.9\kt$ and different adhesion strengths. (a) A relatively small adhesion strength, $\epsilon=3\epsilon_{0}$, leads to a long neck. (b) For $\epsilon=4\epsilon_{0}$ the neck is shorter. (c) For $\epsilon=5\epsilon_{0}$, close to the adhesion strength that leads to membrane rupture, the neck length is comparable to the height of typical membrane fluctuations. The top row of images shows a  side view of system configurations and the bottom row of images gives the corresponding side-view slices.\normalsize}
\label{fig:neck}
\end{figure}

\begin{figure}

\begin{center}
 \includegraphics[width=7.5cm,bb=0 0 812 360]{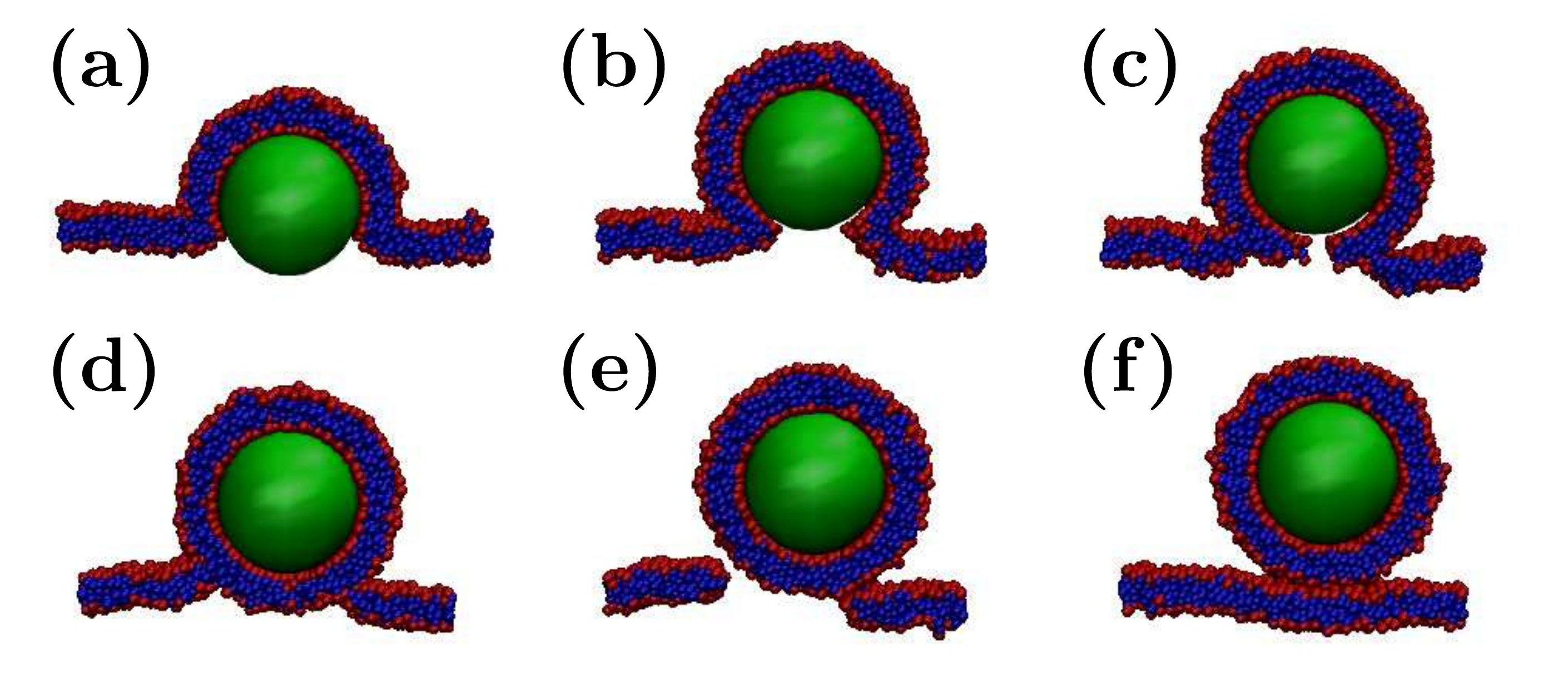}
\end{center}

\caption{\small \textbf{Wrapping via membrane rupture.} Slices of configurations at different times extracted from MD simulations with $\epsilon=5\epsilon_{0}$ ,$R=10\sigma$, and $\kappa=13.9\kt$.
Particle wrapping (a, b) leads to the formation of a pore (c, d). Eventually, the enveloped particle leaves the membrane (e) and the pore closes (f). Configurations shown occured at times (a) $t=2\cdot10^{3}\tau_{0}$,
(b) $t=5\cdot10^{3}\tau_{0}$,(c) $t=5.4\cdot10^{3}\tau_{0}$, (d) $t=5.6\cdot10^{3}\tau_{0}$,(e) $t=8\cdot10^{3}\tau_{0}$, (f) $t=9.5\cdot10^{3}\tau_{0}$.\normalsize}
\label{fig:confbr}
\end{figure}

{\bf Free energy calculations.}
We were particularly interested in the partially wrapped states seen in the dynamical simulations described in the previous section, as the elastic model predicts only  fully wrapped or non-wrapped states. To determine whether or not these observations corresponded to equilibrium configurations,  the free energy was calculated as a function of the penetration using umbrella sampling (section \textbf{System Model}). Calculated free energy projections are shown for three values of $\epsilon$ in Figure 9, 
 for which the finite-time dynamical simulations respectively ended in no wrapping ($\epsilon=0.5\epsilon_0$, Figure 9a), 
 partial wrapping ($\epsilon=\epsilon_0$, Figure 9b), 
 and complete wrapping ($\epsilon=1.25\epsilon_0$, Figure 9c). 
 For the cases of full wrapping and no wrapping, the calculated free energy projections are consistent with the dynamics results. Namely, for $\epsilon=0.5\epsilon_0$ the minimum free energy value corresponds to no wrapping with a steep penalty for increasing  penetration, while for $\epsilon=1.25\epsilon_0$ the free energy decreases monotonically with increasing penetration until the particle is completely wrapped.

\begin{figure}[h!]

\begin{center}
 \includegraphics[width=7.5cm,bb=0 0 860 636]{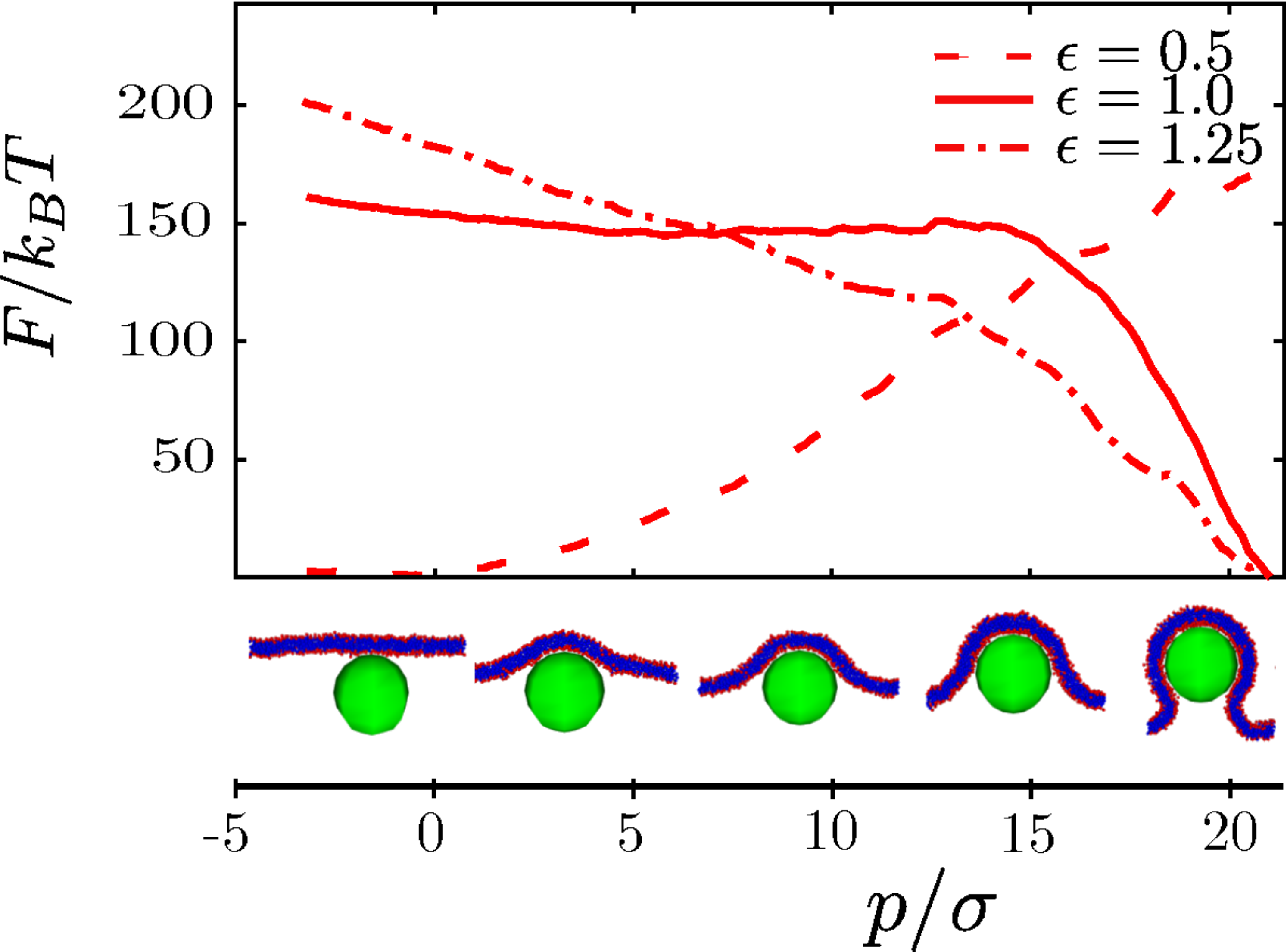}
\end{center}

\caption{\small Free energy profiles as a function of the penetration, \textit{p}, calculated from MD simulations using umbrella sampling are shown  for $R=12\sigma$, $\kappa=13.9\kt$, and indicated values of $\epsilon$. At the bottom, slices of the system as a function of the penetration are shown.\normalsize}
\label{fig:freenR12f}
\end{figure}

In contrast, the minimum value in the free energy profile for $\epsilon=1.0\epsilon_0$ does not correspond to the partially wrapped state observed in the dynamical simulations, but rather corresponds to complete wrapping.  Further comparison of umbrella sampling results to dynamical trajectories suggests that, while  the penetration coordinate $p$ is capable of describing the free energy basins corresponding to the unwrapped and wrapped states, $p$ alone is not sufficient to completely describe the transition dynamics. I.e., $p$ is a suitable order parameter for determining the free energies of the stable states, but not a complete reaction coordinate \cite{Dellago2002}. To see this, we chose a set of configurations from the umbrella sampling trajectories with different values of $p$. For each such configuration we performed several unbiased MD trajectories initialized with velocities using different random number seeds to obtain a crude estimate of the commitment probability \cite{Dellago2002}. We found that trajectories initiated from configurations with small values of $p\lesssim5 \sigma$ fluctuate around that value and configurations with $p\gtrsim18.8\sigma$ progressed steadily to complete wrapping. However, configurations with $5\sigma\leq p    \lesssim 18.8\sigma$ tended to fluctuate around the value of $p$ corresponding to their initial configuration, which is inconsistent with the free energy profile for $p\ge15\sigma$ and indicates the presence of an additional slow degree of freedom at moderate $\epsilon$.

It is not necessary to identify a perfect reaction coordinate to fulfill our primary objective of understanding the phase behavior, but we did attempt to identify the second relevant dynamical degree of freedom. Analysis of umbrella sampling configurations during the equilibration phase of the calculation indicates that, when the particle is held at a fixed penetration, the membrane configuration gradually relaxes to a state of increased adhesion and bending (Figure 10). 
 Thus we expect that a reaction coordinate capable of describing the transition dynamics needs to include an additional collective variable that describes adhesion and/or membrane bending. Investigating this possibility, however, is beyond the scope of the present work focused on the phase behavior.

\begin{figure}[h!]

\begin{center}
 \includegraphics[width=6.0cm,bb=0 0 816 1067]{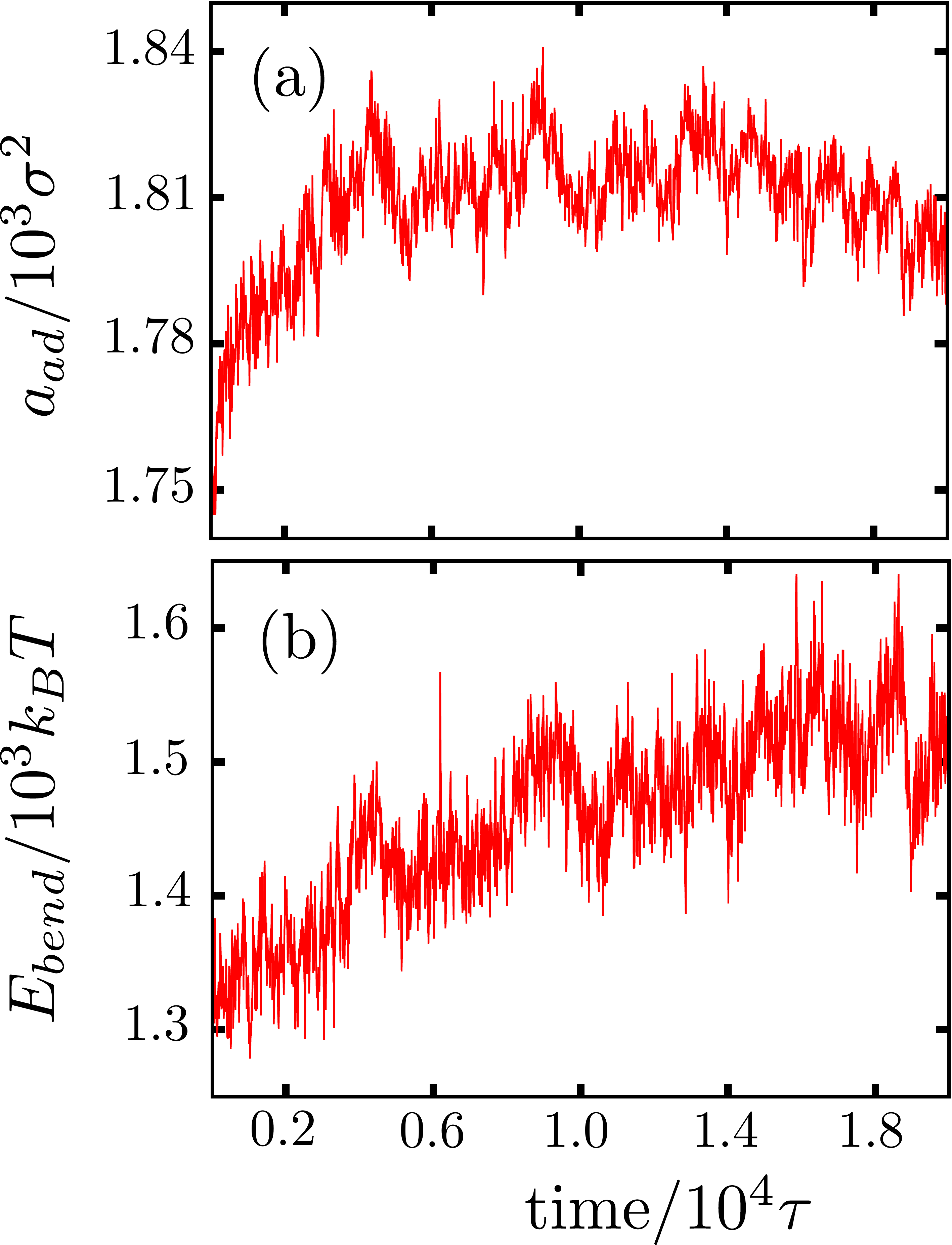}
\end{center}

\caption{\small Evolution of the membrane area in contact with the particle (a) and bending energy (b) as a function of time for MD simulations with a biasing potential holding the penetration near $p=16.9 \sigma$, $U_{\text{bias}}=0.5 \kappa_\text{umb}(p-16.875)^2$, and $\kappa_\text{umb}=200\epsilon_{0}/\sigma^2$. The time courses are averaged over 20 independent trajectories\normalsize}
\label{fig:bendadhevolution}
\end{figure}

\subsection*{Phase diagrams}
\label{ssec:phasediagrams}

Based on the results of dynamical simulations over a wide range of parameters, as well as umbrella sampling at parameter sets near the transition between no wrapping and wrapping, we determined phase diagrams as functions of adhesion energy $\epsilon$, membrane bending modulus $\kappa$, and particle radius $R$ (Figure 12). 
 To enable comparison with the elastic theory Eq.(9), 
 it is essential to note the the theoretical parameter $\epsilon^{*}$ corresponds to the adhesion free energy rather than simply the depth of the head group-particle attractive potential well $\epsilon$. Therefore, we plot the data as a function of the adhesion free energy per area, calculated as $\epsilon^*/\kt=-\eta\log \left[ 1+\int_{s+\sigma}^{\infty}dr \left(  e^{-V_\text{particle-head}(r)}-1 \right) \right]$, Figure 11. 
 Here we have neglected the tiny contribution from  cutting off the potential, we assume that each  lipid head group in contact with the particle approaches roughly along a radial coordinate, and we neglect configurational entropy losses endured by the lipid molecules during adhesion. The parameter sets for which non-wrapping is the equilibrium configuration are shown with * symbols, while the parameter sets which lead to equilibrium wrapping are separated into those which involve long-lived partially wrapped structures (x symbols), complete wrapping (+ symbols), and those for which the membrane undergoes rupture prior to budding ($\Box$ symbols). The wrapping binodal predicted by the elastic theory is shown as a dashed line on each plot. We see that while the theory and simulations agree to within about $0.2\kt$, the theoretical binodal is below the computational results. This discrepancy could occur because we have not accounted for the configurational entropy contributed by the lipids during adhesion or due to the fact that the theory assumes an infinitesimally thin membrane.

\begin{figure}

\begin{center}
 \includegraphics[width=6.0cm,bb=0 0 753 575]{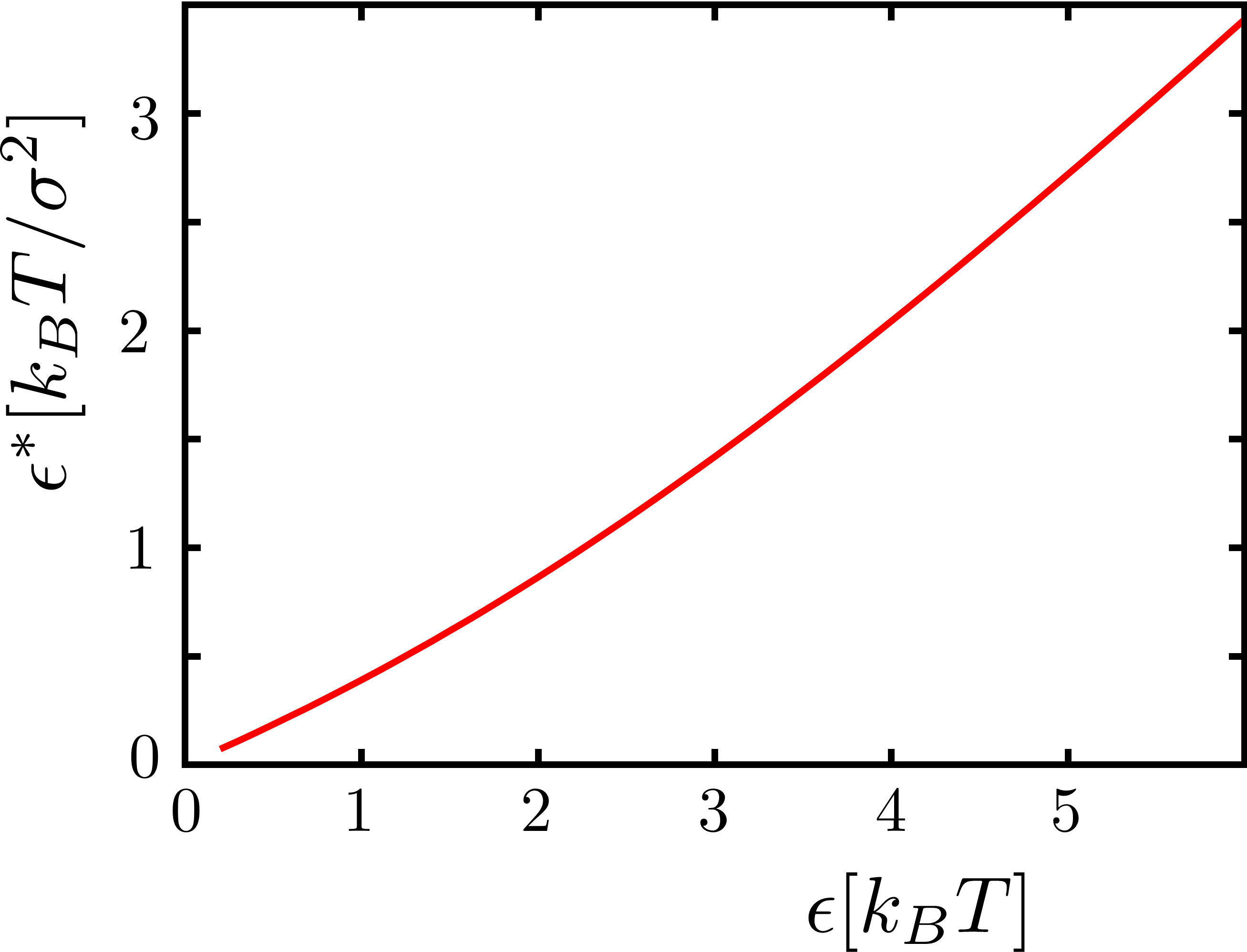}
\end{center}

\caption{\small Relationship between the adhesion energy $\epsilon$ and the adhesion free energy per area $\epsilon*$ for $\kappa=13.9 \kt$.\normalsize}
\label{fig:epsrel}
\end{figure}

The complete phase diagram predicted by the elastic theory is shown in Figure 13. 
 Here, the dashed line is the binodal, given by  $\epsilon*=\frac{2\kappa}{R^{2}}$, below which wrapping is energetically unfavorable and the solid line denotes the spinodal, above which wrapping proceeds without any energetic barrier. In between the lines there is a barrier to wrapping which begins at $p=0$, meaning that there are no long-lived partially wrapped states consistent with this theory for infinite membranes.

\begin{figure}

\begin{center}
 \includegraphics[width=6.5cm,bb=0 0 788 1135]{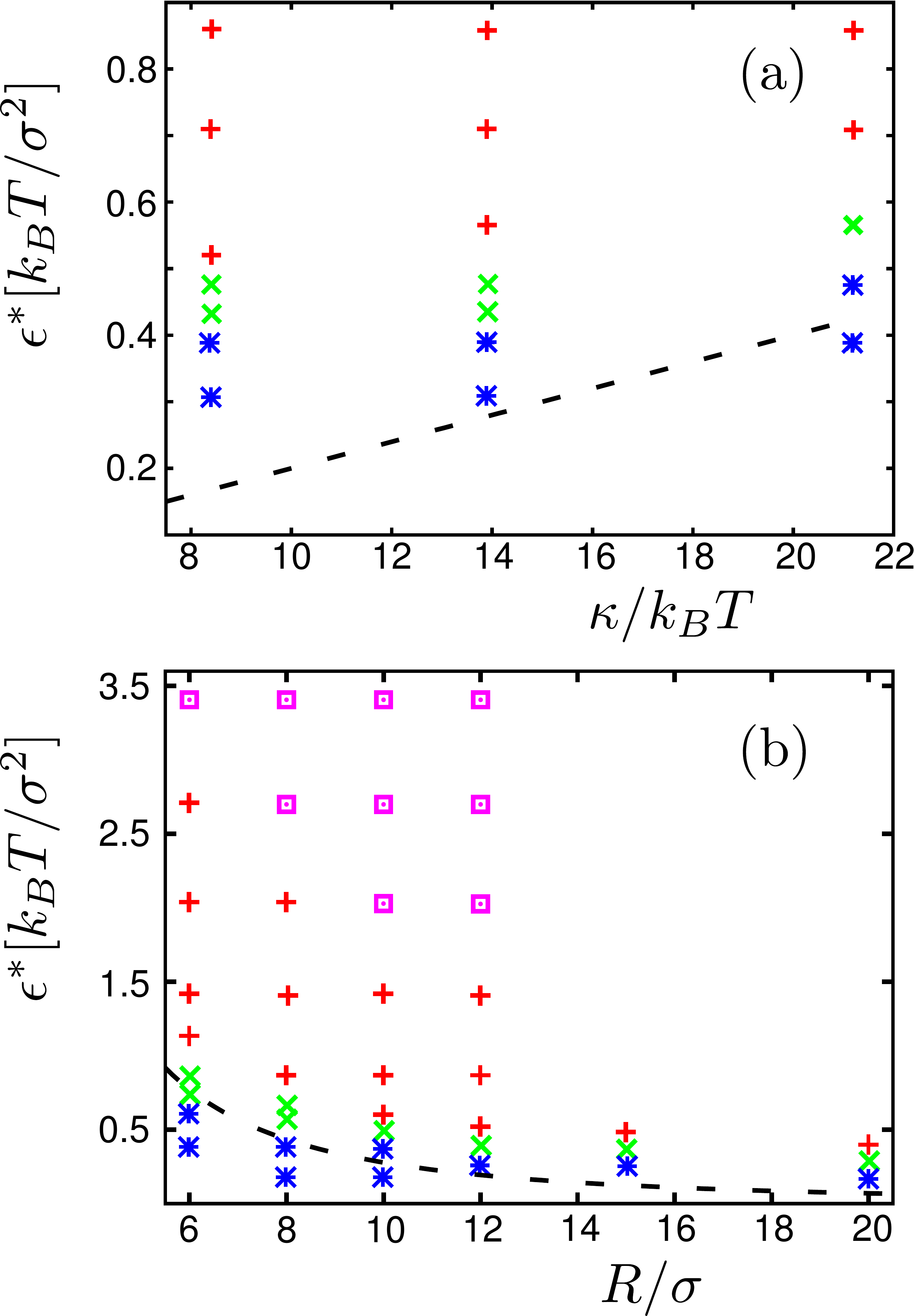}
\end{center}

\caption{\small Phase diagrams obtained from MD simulations and free energy calculations (a) 
 as a function of $\epsilon$ and  $\kappa$ for constant $R=10\sigma$ and (b)   as a function of the particle size $R$ and adhesion free energy $\epsilon*$ for constant bending rigidity, $\kappa=13.9 \kt$. Parameter sets are identified as those which lead to no wrapping ($*$ symbols), long-lived partially wrapped structures (x symbols), complete wrapping (+ symbols), and those for which the membrane undergoes rupture prior to budding ($\Box$ symbols). The wrapping binodal predicted by the elastic theory is shown as a dashed line on each plot. The relationship between the adhesion free energy $\epsilon*$ and the head group-particle attractive well depth $\epsilon$ is given in the text and in Figure 11.\normalsize
}
\label{fig:phasediagsim}
\end{figure}

\begin{figure}
\begin{center}
 \includegraphics[width=6.5cm,bb=0 0 650 1131]{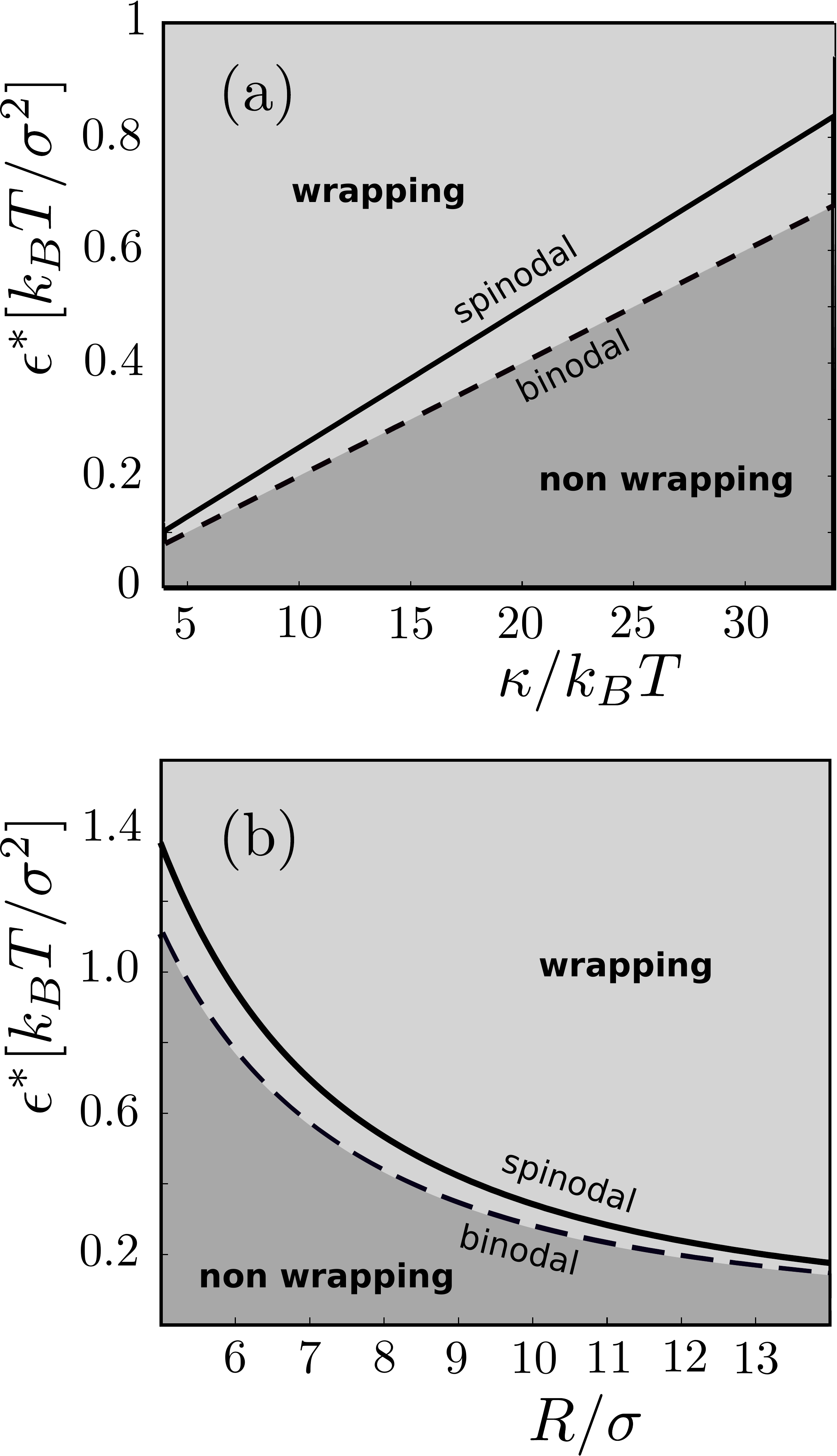}
\end{center}

\caption{\small Phases diagrams obtained from the energy minimization in Eq.(10) 
 (a) 
 as a function of $\epsilon^{*}$ and $\kappa$ for constant $R=10\sigma$ and (b) 
 as a function of the particle size, R,
and the adhesion free energy per area, $\epsilon^{*}$, for constant bending rigidity $\kappa=13.9 \kt$. The binodal ($\epsilon*=\frac{2\kappa}{R^{2}}$) above which wrapping is energetically favorable is indicated by a dashed line and the spinodal, above which wrapping proceeds without an energy barrier, is shown by a solid line. \normalsize}
\label{fig:phasediagtheo}
\end{figure}

\section*{Discussion}
Our dynamical simulations of a minimal molecular model for the process of passive endo- or exocytosis identified four classes of behaviors resulting from the interaction of a particle with a membrane, no wrapping or adhesion resulting in a minimal  perturbation of membrane configurations, partial wrapping, complete wrapping, and wrapping via rupture of the membrane. Equilibrium calculations showed that there are only two equilibrium configurations, corresponding to no wrapping or complete wrapping, and this equilibrium phase behavior for the molecular model  shows strong agreement with the predictions of a simplified elastic theory \cite{Deserno2002}. The primary difference between the elastic model and the finite-time dynamical simulation results is the appearance of long-lived partial wrapping states.

 Since the long-lived partially wrapped states seen in this study could be significant for dynamical, time-sensitive particle uptake processes such as endocytosis or viral budding in living organisms, it is worth comparing them to observations of other models. Most closely related to our results, Yue and Zhang \cite{Yue2011} study a model comparable to ours except that the particle is coated with ligands that experience attractive interactions with membrane lipids. It appears that some of the configurations which they denote as `adhesion' correspond to our long-lived partially wrapped states; however, it would be necessary to perform free energy calculations to determine whether they are metastable as we find here. In contrast to their model, we do not identify any parameter sets for which the particle partially penetrates into the hydrophobic interior of the membrane.
 Partially wrapped states have also been predicted from equilibrium theories in the context of finite-sized systems. Deserno et al. \cite{Deserno2002} examined the budding of colloid from the interior of a spherical vesicle using the same Hamiltonian as introduced in elastic theory described above. They found that partial wrapping corresponds to the equilibrium state when the vesicle size is on the order of the colloid diameter due to the increase in curvature energy of the finite-sized vesicle.  We similarly obtain partial wrapping configurations as equilibrium solutions if we introduce finite size into the elastic theory studied here (Eq.(9)) 
 by minimizing the energy of the elastic theory (Eq.(10)) 
 with the total membrane surface area constrained to $16R^2$, so that the wrapping area plus the rim area cannot
exceed the total area.  Zhang and Nguyen \cite{Zhang2008} also identify partially wrapped states as equilibrium solutions to an elastic theory, but they observed that the catenoid configuration is the only solution to the full variational problem for a tensionless infinite membrane. This solution implies that the elastic energy of the rim is always zero, and thus wrapping is only determined by the balance between the
bending energy in the wrapped region and the adhesion energy, which does not lead to partial wrapping states. Because the toroid approximation for membrane configurations assumed in our elastic model is more restrictive than the full variational problem, the wrapping binodal shown in Figure 13 
 is shifted to slightly higher values of the adhesion energy $\epsilon$ than obtained for their theory \cite{Zhang2008}, but the behavior is qualitatively unchanged. Importantly, neither theory predicts the partially wrapped state as a metastable configuration in an infinite membrane. 

The membrane size in our simulations was chosen large enough to ensure that the theoretically predicted finite-size effects would not affect our results. To confirm that this was the case, we ran additional dynamical simulations and umbrella sampling calculations with membranes which were $50 \%$ and $100\%$ larger (16200 and 28800 lipids respectively). The simulation results were the same for all three membrane sizes.

Finally, we consider our minimal model for passive endo- or exocytosis in the context of physical systems. Based on the length scales discussed in section \textbf{Parameters}, the particle diameters
in our simulations range from 9 to 36 nm. Nanoparticles are available in a wide range of sizes, with particles smaller than 50 nm undergoing the most efficient uptake \cite{Jiang2008}, and our simulation results indicate a range of possible uptake pathways. Our simulated particles are somewhat smaller than the size of viral capsids that undergo budding, which range from about 40 nm (e.g. hepadnavirus \cite{Seeger2000}) to hundreds of nanometers, but the results can be extrapolated into that range. As shown in Figure 13 
 the adhesion-wrapping transition decreases with radius as $1/R$ for constant bending rigidity.

Our simulation results indicate that the existence of attractive interactions between a particle and lipid head groups, which has been proposed as the minimal requirement for viral budding \cite{Garoff1998,Garoff2004}, is indeed sufficient to drive efficient wrapping. However, to  avoid stalled partially-wrapping states and membrane rupture, the system would be confined to a relatively narrow range of adhesion strengths spanning about $2 \kt/\sigma^2$ (the + symbols in Figure 12). 
 While this result is qualitative, since the range increases in width with the particle size and the three-bead representation of the lipid molecule may lead to model membranes which are more susceptible to rupture than those comprised of a more realistic lipid, it does establish important constraints on viral evolution if budding is limited to these ingredients. However, depending on the viral system, a number of additional phenomena contribute to budding, including membrane-associated viral envelope or spike proteins, preferential budding from lipid rafts \cite{Ono2001}, the use of cytoskeletal machinery to actively drive or assist budding \cite{Gladnikoff2009,Taylor2011} or scission \cite{Demirov2004,Baumgartel2011}, and the ability of the virus to remodel cell membrane properties\cite{Chan2010}. These effects can broaden the range of functional adhesion energies; e.g. using actin to drive assembly and budding \cite{Gladnikoff2009} could enable efficient viral egress even for adhesion energies at which spontaneous dynamics become stalled. In this case the presence of a barrier to budding could serve as a regulatory feature. The agreement between our simulation results and the elastic theory over some ranges of parameter space indicates that some or all of these effects could in principle be captured by extending existing elastic theories along the lines of Liu et. al.'s description of active endocytosis \cite{Liu2009}, but care would be required to include all relevant slow degrees of freedom near  transitions between wrapping and no wrapping.



\providecommand*\mcitethebibliography{\thebibliography}
\csname @ifundefined\endcsname{endmcitethebibliography}
  {\let\endmcitethebibliography\endthebibliography}{}


\begin{mcitethebibliography}{46}
\providecommand*\natexlab[1]{#1}
\providecommand*\mciteSetBstSublistMode[1]{}
\providecommand*\mciteSetBstMaxWidthForm[2]{}
\providecommand*\mciteBstWouldAddEndPuncttrue
  {\def\EndOfBibitem{\unskip.}}
\providecommand*\mciteBstWouldAddEndPunctfalse
  {\let\EndOfBibitem\relax}
\providecommand*\mciteSetBstMidEndSepPunct[3]{}
\providecommand*\mciteSetBstSublistLabelBeginEnd[3]{}
\providecommand*\EndOfBibitem{}
\mciteSetBstSublistMode{f}
\mciteSetBstMaxWidthForm{subitem}{(\alph{mcitesubitemcount})}
\mciteSetBstSublistLabelBeginEnd
  {\mcitemaxwidthsubitemform\space}
  {\relax}
  {\relax}

\bibitem[Nel et~al.(2009)Nel, M\"{a}dler, Velegol, Xia, Hoek, Somasundaran,
  Klaessig, Castranova, and Thompson]{Nel2009}
Nel,~A.~E.; M\"{a}dler,~L.; Velegol,~D.; Xia,~T.; Hoek,~E. M.~V.;
  Somasundaran,~P.; Klaessig,~F.; Castranova,~V.; Thompson,~M. \emph{Nat Mater}
  \textbf{2009}, \emph{8}, 543--557\relax
\mciteBstWouldAddEndPuncttrue
\mciteSetBstMidEndSepPunct{\mcitedefaultmidpunct}
{\mcitedefaultendpunct}{\mcitedefaultseppunct}\relax
\EndOfBibitem
\bibitem[Mitragotri and Lahann(2009)Mitragotri, and Lahann]{Mitragotri2009}
Mitragotri,~S.; Lahann,~J. \emph{Nat Mater} \textbf{2009}, \emph{8},
  15--23\relax
\mciteBstWouldAddEndPuncttrue
\mciteSetBstMidEndSepPunct{\mcitedefaultmidpunct}
{\mcitedefaultendpunct}{\mcitedefaultseppunct}\relax
\EndOfBibitem
\bibitem[Poland et~al.(2008)Poland, Duffin, Kinloch, Maynard, Wallace, Seaton,
  Stone, Brown, Macnee, and Donaldson]{Poland2008}
Poland,~C.~A.; Duffin,~R.; Kinloch,~I.; Maynard,~A.; Wallace,~W. A.~H.;
  Seaton,~A.; Stone,~V.; Brown,~S.; Macnee,~W.; Donaldson,~K. \emph{Nat
  Nanotechnol} \textbf{2008}, \emph{3}, 423--428\relax
\mciteBstWouldAddEndPuncttrue
\mciteSetBstMidEndSepPunct{\mcitedefaultmidpunct}
{\mcitedefaultendpunct}{\mcitedefaultseppunct}\relax
\EndOfBibitem
\bibitem[Liu et~al.(2009)Liu, Sun, Drubin, and Oster]{Liu2009}
Liu,~J.; Sun,~Y.; Drubin,~D.~G.; Oster,~G.~F. \emph{PLoS Biol} \textbf{2009},
  \emph{7}, e1000204\relax
\mciteBstWouldAddEndPuncttrue
\mciteSetBstMidEndSepPunct{\mcitedefaultmidpunct}
{\mcitedefaultendpunct}{\mcitedefaultseppunct}\relax
\EndOfBibitem
\bibitem[Welsch et~al.(2007)Welsch, M\"{u}ller, and Kr\"{a}usslich]{Welsch2007}
Welsch,~S.; M\"{u}ller,~B.; Kr\"{a}usslich,~H.-G. \emph{FEBS Letters}
  \textbf{2007}, \emph{581}, 2089 -- 2097, <ce:title>Membrane
  Trafficking</ce:title>\relax
\mciteBstWouldAddEndPuncttrue
\mciteSetBstMidEndSepPunct{\mcitedefaultmidpunct}
{\mcitedefaultendpunct}{\mcitedefaultseppunct}\relax
\EndOfBibitem
\bibitem[Gladnikoff et~al.(2009)Gladnikoff, Shimoni, Gov, and
  Rousso]{Gladnikoff2009}
Gladnikoff,~M.; Shimoni,~E.; Gov,~N.~S.; Rousso,~I. \emph{Biophysical Journal}
  \textbf{2009}, \emph{97}, 2419--2428\relax
\mciteBstWouldAddEndPuncttrue
\mciteSetBstMidEndSepPunct{\mcitedefaultmidpunct}
{\mcitedefaultendpunct}{\mcitedefaultseppunct}\relax
\EndOfBibitem
\bibitem[Baumgartel et~al.(2011)Baumgartel, Ivanchenko, Dupont, Sergeev,
  Wiseman, Krausslich, Brauchle, Muller, and Lamb]{Baumgartel2011}
Baumgartel,~V.; Ivanchenko,~S.; Dupont,~A.; Sergeev,~M.; Wiseman,~P.~W.;
  Krausslich,~H.-G.; Brauchle,~C.; Muller,~B.; Lamb,~D.~C. \emph{Nat Cell Biol}
  \textbf{2011}, \emph{13}, 469--474\relax
\mciteBstWouldAddEndPuncttrue
\mciteSetBstMidEndSepPunct{\mcitedefaultmidpunct}
{\mcitedefaultendpunct}{\mcitedefaultseppunct}\relax
\EndOfBibitem
\bibitem[Helenius et~al.(1977)Helenius, Fries, and Kartenbeck]{Helenius1977}
Helenius,~A.; Fries,~E.; Kartenbeck,~J. \emph{J Cell Biol} \textbf{1977},
  \emph{75}, 866--880\relax
\mciteBstWouldAddEndPuncttrue
\mciteSetBstMidEndSepPunct{\mcitedefaultmidpunct}
{\mcitedefaultendpunct}{\mcitedefaultseppunct}\relax
\EndOfBibitem
\bibitem[von Bonsdorff and Harrison(1978)von Bonsdorff, and
  Harrison]{Bonsdorff1978}
von Bonsdorff,~C.~H.; Harrison,~S.~C. \emph{Journal of Virology} \textbf{1978},
  \emph{28}, 578--583\relax
\mciteBstWouldAddEndPuncttrue
\mciteSetBstMidEndSepPunct{\mcitedefaultmidpunct}
{\mcitedefaultendpunct}{\mcitedefaultseppunct}\relax
\EndOfBibitem
\bibitem[Bihan et~al.(2009)Bihan, Bonnafous, Marak, Bickel, Tr\'{e}pout,
  Mornet, Haas, Talbot, Taveau, and Lambert]{Bihan2009}
Bihan,~O.~L.; Bonnafous,~P.; Marak,~L.; Bickel,~T.; Tr\'{e}pout,~S.;
  Mornet,~S.; Haas,~F.~D.; Talbot,~H.; Taveau,~J.-C.; Lambert,~O. \emph{J
  Struct Biol} \textbf{2009}, \emph{168}, 419--425\relax
\mciteBstWouldAddEndPuncttrue
\mciteSetBstMidEndSepPunct{\mcitedefaultmidpunct}
{\mcitedefaultendpunct}{\mcitedefaultseppunct}\relax
\EndOfBibitem
\bibitem[Hurley et~al.(2010)Hurley, Boura, Carlson, and
  R\'{o}zycki]{Hurley2010}
Hurley,~J.~H.; Boura,~E.; Carlson,~L.-A.; R\'{o}zycki,~B. \emph{Cell}
  \textbf{2010}, \emph{143}, 875 -- 887\relax
\mciteBstWouldAddEndPuncttrue
\mciteSetBstMidEndSepPunct{\mcitedefaultmidpunct}
{\mcitedefaultendpunct}{\mcitedefaultseppunct}\relax
\EndOfBibitem
\bibitem[Solon et~al.(2005)Solon, Gareil, Bassereau, and Gaudin]{Solon2005}
Solon,~J.; Gareil,~O.; Bassereau,~P.; Gaudin,~Y. \emph{J Gen Virol}
  \textbf{2005}, \emph{86}, 3357--3363\relax
\mciteBstWouldAddEndPuncttrue
\mciteSetBstMidEndSepPunct{\mcitedefaultmidpunct}
{\mcitedefaultendpunct}{\mcitedefaultseppunct}\relax
\EndOfBibitem
\bibitem[Popova et~al.(2010)Popova, Popov, and G\"{o}ttlinger]{Popova2010b}
Popova,~E.; Popov,~S.; G\"{o}ttlinger,~H.~G. \emph{J Virol} \textbf{2010},
  \emph{84}, 6590--6597\relax
\mciteBstWouldAddEndPuncttrue
\mciteSetBstMidEndSepPunct{\mcitedefaultmidpunct}
{\mcitedefaultendpunct}{\mcitedefaultseppunct}\relax
\EndOfBibitem
\bibitem[Rossman et~al.(2010)Rossman, Jing, Leser, and Lamb]{Rossman2010}
Rossman,~J.~S.; Jing,~X.; Leser,~G.~P.; Lamb,~R.~A. \emph{Cell} \textbf{2010},
  \emph{142}, 902--913\relax
\mciteBstWouldAddEndPuncttrue
\mciteSetBstMidEndSepPunct{\mcitedefaultmidpunct}
{\mcitedefaultendpunct}{\mcitedefaultseppunct}\relax
\EndOfBibitem
\bibitem[Deserno and Gelbart(2002)Deserno, and Gelbart]{Deserno2002}
Deserno,~M.; Gelbart,~W.~M. \emph{The Journal of Physical Chemistry B}
  \textbf{2002}, \emph{106}, 5543--5552\relax
\mciteBstWouldAddEndPuncttrue
\mciteSetBstMidEndSepPunct{\mcitedefaultmidpunct}
{\mcitedefaultendpunct}{\mcitedefaultseppunct}\relax
\EndOfBibitem
\bibitem[Gao et~al.(2005)Gao, Shi, and Freund]{Gao2005}
Gao,~H.~J.; Shi,~W.~D.; Freund,~L.~B. \emph{Proceedings of the National Academy
  of Sciences of the United States of America} \textbf{2005}, \emph{102},
  9469--9474\relax
\mciteBstWouldAddEndPuncttrue
\mciteSetBstMidEndSepPunct{\mcitedefaultmidpunct}
{\mcitedefaultendpunct}{\mcitedefaultseppunct}\relax
\EndOfBibitem
\bibitem[Zhang and Nguyen(2008)Zhang, and Nguyen]{Zhang2008}
Zhang,~R.; Nguyen,~T.~T. \emph{Phys. Rev. E} \textbf{2008}, \emph{78},
  051903\relax
\mciteBstWouldAddEndPuncttrue
\mciteSetBstMidEndSepPunct{\mcitedefaultmidpunct}
{\mcitedefaultendpunct}{\mcitedefaultseppunct}\relax
\EndOfBibitem
\bibitem[Fo\v{s}nari\v{c} et~al.(2009)Fo\v{s}nari\v{c}, Igli\v{c}, Kroll, and
  May]{Fosnaric2009}
Fo\v{s}nari\v{c},~M.; Igli\v{c},~A.; Kroll,~D.~M.; May,~S. \emph{The Journal of
  Chemical Physics} \textbf{2009}, \emph{131}, 105103\relax
\mciteBstWouldAddEndPuncttrue
\mciteSetBstMidEndSepPunct{\mcitedefaultmidpunct}
{\mcitedefaultendpunct}{\mcitedefaultseppunct}\relax
\EndOfBibitem
\bibitem[Li and Gu(2010)Li, and Gu]{Li2010}
Li,~Y.; Gu,~N. \emph{Journal of Physical Chemistry B} \textbf{2010},
  \emph{114}, 2749--2754\relax
\mciteBstWouldAddEndPuncttrue
\mciteSetBstMidEndSepPunct{\mcitedefaultmidpunct}
{\mcitedefaultendpunct}{\mcitedefaultseppunct}\relax
\EndOfBibitem
\bibitem[Ginzburg and Balijepalli(2007)Ginzburg, and Balijepalli]{Ginzburg2007}
Ginzburg,~V.~V.; Balijepalli,~S. \emph{Nano Letters} \textbf{2007}, \emph{7},
  3716--3722\relax
\mciteBstWouldAddEndPuncttrue
\mciteSetBstMidEndSepPunct{\mcitedefaultmidpunct}
{\mcitedefaultendpunct}{\mcitedefaultseppunct}\relax
\EndOfBibitem
\bibitem[Smith et~al.(2007)Smith, Jasnow, and Balazs]{Smith2007}
Smith,~K.~A.; Jasnow,~D.; Balazs,~A.~C. \emph{The Journal of Chemical Physics}
  \textbf{2007}, \emph{127}, 084703\relax
\mciteBstWouldAddEndPuncttrue
\mciteSetBstMidEndSepPunct{\mcitedefaultmidpunct}
{\mcitedefaultendpunct}{\mcitedefaultseppunct}\relax
\EndOfBibitem
\bibitem[Yue and Zhang(2011)Yue, and Zhang]{Yue2011}
Yue,~T.~T.; Zhang,~X.~R. \emph{Soft Matter} \textbf{2011}, \emph{7},
  9104--9112\relax
\mciteBstWouldAddEndPuncttrue
\mciteSetBstMidEndSepPunct{\mcitedefaultmidpunct}
{\mcitedefaultendpunct}{\mcitedefaultseppunct}\relax
\EndOfBibitem
\bibitem[Yang and Ma(2011)Yang, and Ma]{Yang2011b}
Yang,~K.; Ma,~Y.~Q. \emph{Australian Journal of Chemistry} \textbf{2011},
  \emph{64}, 894--899\relax
\mciteBstWouldAddEndPuncttrue
\mciteSetBstMidEndSepPunct{\mcitedefaultmidpunct}
{\mcitedefaultendpunct}{\mcitedefaultseppunct}\relax
\EndOfBibitem
\bibitem[Vacha et~al.(2011)Vacha, Martinez-Veracoechea, and Frenkel]{Vacha2011}
Vacha,~R.; Martinez-Veracoechea,~F.~J.; Frenkel,~D. \emph{Nano Letters}
  \textbf{2011}, \emph{11}, 5391--5395\relax
\mciteBstWouldAddEndPuncttrue
\mciteSetBstMidEndSepPunct{\mcitedefaultmidpunct}
{\mcitedefaultendpunct}{\mcitedefaultseppunct}\relax
\EndOfBibitem
\bibitem[Garoff et~al.(1998)Garoff, Hewson, and Opstelten]{Garoff1998}
Garoff,~H.; Hewson,~R.; Opstelten,~D.~J. \emph{Microbiol Mol Biol Rev}
  \textbf{1998}, \emph{62}, 1171--1190\relax
\mciteBstWouldAddEndPuncttrue
\mciteSetBstMidEndSepPunct{\mcitedefaultmidpunct}
{\mcitedefaultendpunct}{\mcitedefaultseppunct}\relax
\EndOfBibitem
\bibitem[Demirov and Freed(2004)Demirov, and Freed]{Demirov2004}
Demirov,~D.~G.; Freed,~E.~O. \emph{Virus Res} \textbf{2004}, \emph{106},
  87--102\relax
\mciteBstWouldAddEndPuncttrue
\mciteSetBstMidEndSepPunct{\mcitedefaultmidpunct}
{\mcitedefaultendpunct}{\mcitedefaultseppunct}\relax
\EndOfBibitem
\bibitem[Chan et~al.(2010)Chan, Tanner, and Wenk]{Chan2010}
Chan,~R.~B.; Tanner,~L.; Wenk,~M.~R. \emph{Chem Phys Lipids} \textbf{2010},
  \emph{163}, 449--459\relax
\mciteBstWouldAddEndPuncttrue
\mciteSetBstMidEndSepPunct{\mcitedefaultmidpunct}
{\mcitedefaultendpunct}{\mcitedefaultseppunct}\relax
\EndOfBibitem
\bibitem[Cooke et~al.(2005)Cooke, Kremer, and Deserno]{Cooke2005}
Cooke,~I.~R.; Kremer,~K.; Deserno,~M. \emph{Phys. Rev. E} \textbf{2005},
  \emph{72}, 011506\relax
\mciteBstWouldAddEndPuncttrue
\mciteSetBstMidEndSepPunct{\mcitedefaultmidpunct}
{\mcitedefaultendpunct}{\mcitedefaultseppunct}\relax
\EndOfBibitem
\bibitem[Weeks et~al.(1971)Weeks, Chandler, and Andersen]{Weeks1971}
Weeks,~J.~D.; Chandler,~D.; Andersen,~H.~C. \emph{J. Chem. Phys.}
  \textbf{1971}, \emph{54}, 5237+\relax
\mciteBstWouldAddEndPuncttrue
\mciteSetBstMidEndSepPunct{\mcitedefaultmidpunct}
{\mcitedefaultendpunct}{\mcitedefaultseppunct}\relax
\EndOfBibitem
\bibitem[Grest and Kremer(1986)Grest, and Kremer]{Grest1986}
Grest,~G.~S.; Kremer,~K. \emph{Phys. Rev. A} \textbf{1986}, \emph{33},
  3628--3631\relax
\mciteBstWouldAddEndPuncttrue
\mciteSetBstMidEndSepPunct{\mcitedefaultmidpunct}
{\mcitedefaultendpunct}{\mcitedefaultseppunct}\relax
\EndOfBibitem
\bibitem[Garoff et~al.(2004)Garoff, Sj\"{o}berg, and Cheng]{Garoff2004}
Garoff,~H.; Sj\"{o}berg,~M.; Cheng,~R.~H. \emph{Virus Res} \textbf{2004},
  \emph{106}, 103--116\relax
\mciteBstWouldAddEndPuncttrue
\mciteSetBstMidEndSepPunct{\mcitedefaultmidpunct}
{\mcitedefaultendpunct}{\mcitedefaultseppunct}\relax
\EndOfBibitem
\bibitem[Frenkel and Smit(2002)Frenkel, and Smit]{Frenkel2002a}
Frenkel,~D.; Smit,~B. \emph{Understanding molecular simulation: from algorithms
  to applications}, 2nd ed.; Academic: San Diego, Calif. ; London, 2002\relax
\mciteBstWouldAddEndPuncttrue
\mciteSetBstMidEndSepPunct{\mcitedefaultmidpunct}
{\mcitedefaultendpunct}{\mcitedefaultseppunct}\relax
\EndOfBibitem
\bibitem[Kolb and D\"{u}nweg(1999)Kolb, and D\"{u}nweg]{Kolb1999}
Kolb,~A.; D\"{u}nweg,~B. \emph{The Journal of Chemical Physics} \textbf{1999},
  \emph{111}, 4453--4459\relax
\mciteBstWouldAddEndPuncttrue
\mciteSetBstMidEndSepPunct{\mcitedefaultmidpunct}
{\mcitedefaultendpunct}{\mcitedefaultseppunct}\relax
\EndOfBibitem
\bibitem[Torrie and Valleau(1977)Torrie, and Valleau]{Torrie1977}
Torrie,~G.~M.; Valleau,~J.~P. \emph{J Comput Phys} \textbf{1977}, \emph{23},
  187--199\relax
\mciteBstWouldAddEndPuncttrue
\mciteSetBstMidEndSepPunct{\mcitedefaultmidpunct}
{\mcitedefaultendpunct}{\mcitedefaultseppunct}\relax
\EndOfBibitem
\bibitem[Kumar et~al.(1992)Kumar, Bouzida, Swendsen, Kollman, and
  Rosenberg]{Kumar1992}
Kumar,~S.; Bouzida,~D.; Swendsen,~R.~H.; Kollman,~P.~A.; Rosenberg,~J.~M.
  \emph{J. Comput. Chem.} \textbf{1992}, \emph{13}, 1011--1021\relax
\mciteBstWouldAddEndPuncttrue
\mciteSetBstMidEndSepPunct{\mcitedefaultmidpunct}
{\mcitedefaultendpunct}{\mcitedefaultseppunct}\relax
\EndOfBibitem
\bibitem[Grossfield()]{WHAM}
Grossfield,~A. WHAM: the weighted histogram analysis method, version 2.0.6.
\url{http://membrane.urmc.rochester.edu/content/wham}\relax
\mciteBstWouldAddEndPuncttrue
\mciteSetBstMidEndSepPunct{\mcitedefaultmidpunct}
{\mcitedefaultendpunct}{\mcitedefaultseppunct}\relax
\EndOfBibitem
\bibitem[Helfrich(1973)]{Helfrich1973}
Helfrich,~W. \emph{Zeitschrift Fur Naturforschung C-a Journal of Biosciences}
  \textbf{1973}, \emph{C 28}, 693--703\relax
\mciteBstWouldAddEndPuncttrue
\mciteSetBstMidEndSepPunct{\mcitedefaultmidpunct}
{\mcitedefaultendpunct}{\mcitedefaultseppunct}\relax
\EndOfBibitem
\bibitem[do~Carmo(1976)]{DoCarmo1976}
do~Carmo,~M.~P. \emph{Differential Geometry of Curves and Surfaces}; Prentice
  Hall, 1976\relax
\mciteBstWouldAddEndPuncttrue
\mciteSetBstMidEndSepPunct{\mcitedefaultmidpunct}
{\mcitedefaultendpunct}{\mcitedefaultseppunct}\relax
\EndOfBibitem
\bibitem[Li et~al.(2008)Li, Chen, and Gu]{Li2008}
Li,~Y.; Chen,~X.; Gu,~N. \emph{The Journal of Physical Chemistry B}
  \textbf{2008}, \emph{112}, 16647--16653\relax
\mciteBstWouldAddEndPuncttrue
\mciteSetBstMidEndSepPunct{\mcitedefaultmidpunct}
{\mcitedefaultendpunct}{\mcitedefaultseppunct}\relax
\EndOfBibitem
\bibitem[Humphrey et~al.(1996)Humphrey, Dalke, and Schulten]{Humphrey1996}
Humphrey,~W.; Dalke,~A.; Schulten,~K. \emph{J. Mol. Graph.} \textbf{1996},
  \emph{14}, 33--38\relax
\mciteBstWouldAddEndPuncttrue
\mciteSetBstMidEndSepPunct{\mcitedefaultmidpunct}
{\mcitedefaultendpunct}{\mcitedefaultseppunct}\relax
\EndOfBibitem
\bibitem[Dellago et~al.(2002)Dellago, Bolhuis, and Geissler]{Dellago2002}
Dellago,~C.; Bolhuis,~P.~G.; Geissler,~P.~L. \emph{Advances in Chemical
  Physics, Vol 123}; Advances in Chemical Physics; 2002; Vol. 123; pp
  1--78\relax
\mciteBstWouldAddEndPuncttrue
\mciteSetBstMidEndSepPunct{\mcitedefaultmidpunct}
{\mcitedefaultendpunct}{\mcitedefaultseppunct}\relax
\EndOfBibitem
\bibitem[Jiang et~al.(2008)Jiang, Kim, Rutka, and Chan]{Jiang2008}
Jiang,~W.; Kim,~B. Y.~S.; Rutka,~J.~T.; Chan,~W. C.~W. \emph{Nat Nanotechnol}
  \textbf{2008}, \emph{3}, 145--150\relax
\mciteBstWouldAddEndPuncttrue
\mciteSetBstMidEndSepPunct{\mcitedefaultmidpunct}
{\mcitedefaultendpunct}{\mcitedefaultseppunct}\relax
\EndOfBibitem
\bibitem[Seeger and Mason(2000)Seeger, and Mason]{Seeger2000}
Seeger,~C.; Mason,~W.~S. \emph{Microbiol Mol Biol Rev} \textbf{2000},
  \emph{64}, 51--68\relax
\mciteBstWouldAddEndPuncttrue
\mciteSetBstMidEndSepPunct{\mcitedefaultmidpunct}
{\mcitedefaultendpunct}{\mcitedefaultseppunct}\relax
\EndOfBibitem
\bibitem[Ono and Freed(2001)Ono, and Freed]{Ono2001}
Ono,~A.; Freed,~E.~O. \emph{Proc Natl Acad Sci U S A} \textbf{2001}, \emph{98},
  13925--13930\relax
\mciteBstWouldAddEndPuncttrue
\mciteSetBstMidEndSepPunct{\mcitedefaultmidpunct}
{\mcitedefaultendpunct}{\mcitedefaultseppunct}\relax
\EndOfBibitem
\bibitem[Taylor et~al.(2011)Taylor, Koyuncu, and Enquist]{Taylor2011}
Taylor,~M.~P.; Koyuncu,~O.~O.; Enquist,~L.~W. \emph{Nat Rev Microbiol}
  \textbf{2011}, \emph{9}, 427--439\relax
\mciteBstWouldAddEndPuncttrue
\mciteSetBstMidEndSepPunct{\mcitedefaultmidpunct}
{\mcitedefaultendpunct}{\mcitedefaultseppunct}\relax
\EndOfBibitem
\end{mcitethebibliography}
\end{document}